\documentclass{lmcs}
\pdfoutput=1

\usepackage{lastpage}
\lmcsdoi{15}{1}{10}
\lmcsheading{}{\pageref{LastPage}}{}{}%
{Jul.~17,~2017}{Feb.~06,~2019}{}

\usepackage{hyperref}
\usepackage{amssymb}
\def\jlb#1{\relax}
\def\aa#1{\relax}

\def\q{\hbox to 2em{\hfil}\ignorespaces}
\def\g{\gamma}
\def\l{\lambda}

\def\D{\mathcal{D}}
\def\B{\mathcal{B}}

\def\dto{\Rightarrow}
\def\cls#1{\overline{#1}}

\begin{document}

\title[Relative Entailment Among Probabilistic Implications]{Relative Entailment Among Probabilistic Implications}
\titlecomment{This journal paper is an archival, nontrivially extended version 
of the conference paper \cite{AtsBalLICS}, where we 
prove relevant particular cases of the results presented here,
within a more focused view, due to the timing and to the page limits.}

\author{Albert Atserias}
\address{Department of Computer Science,
Universitat Polit\`ecnica de Catalunya}
\thanks{First author partially supported by European Research Council (ERC) under
the European Union's Horizon 2020 research and innovation programme,
grant agreement ERC-2014-CoG 648276 (AUTAR), and MINECO project
TIN2013-48031-C4-1-P (TASSAT2).}
\author{Jos\'e L. Balc\'azar}
\thanks{Second author partially supported by
SGR2014-890 (MACDA) of the Generalitat de Catalunya and MINECO 
project TIN2014-57226-P (APCOM)}
\author{Marie Ely Piceno}
\thanks{Third author supported by Conacyt, M\'exico, and MINECO 
project TIN2014-57226-P (APCOM)}


\begin{abstract}
\noindent
We study a natural 
variant of the
implicational fragment of propositional logic.
Its formulas are pairs of conjunctions of positive literals, 
related together by an im\-pli\-ca\-tio\-nal-like connective; the 
semantics of this sort of implication 
is defined in terms of a threshold on a conditional probability 
of the consequent, given the antecedent: we are dealing with 
what the data analysis community calls confidence of partial 
implications or association rules. Existing studies of redundancy 
among these partial implications
have characterized so far only 
entailment from one premise and 
entailment from two premises,
both in the stand-alone case and in the case of presence
of additional classical implications
(this is what we call ``relative entailment'').
By exploiting a previously noted alternative view of the entailment
in terms of linear programming duality, we characterize exactly the
cases of entailment from arbitrary numbers of premises,
again both in the stand-alone case and in the case of presence
of additional classical implications. As a result, 
we obtain decision algorithms of better complexity;
additionally, for each potential case of entailment, 
we identify a
critical confidence threshold 
and show that it
is, actually, intrinsic to each
set of premises and antecedent of the conclusion.
\end{abstract}

\maketitle

\section{Introduction}


The quite deep issue of how to represent human knowledge
in a way that is most useful for applications has been
present in research for decades now. 
Often, knowledge representation is necessary in a context of
incomplete information, whereby inductive processes are 
required in addition.
As a result, two facets that are common to a great number of works in 
knowledge representation, and particularly more so in 
contexts of inductive inference, machine learning, or
data analysis, are logic and probability.

Adding probability-based mechanisms to already expressive 
logics enhances their expressiveness and usefulness, but pays 
heavy prices in terms of computational difficulty. 
Even without probability, certain degrees of expressivity
and computational feasibility are known to be incompatible,
and this is reflected in the undecidability results for many logics.
In other cases, the balance between expressivity and feasibility
hinges on often open complexity-theoretic statements. 
To work only within logics known to be polynomially tractable 
may imply serious expressiveness limitations. Premier examples
of polynomially tractable cases are Horn logics.

Literally hundreds of studies have explored
this difficult balance. Already within the limits of
the machine learning perspective, we could mention a 
large number of references such as those cited in the book
\cite{deRaedt}; as well as existing studies, like \cite{GUHA}, 
that include fragments that relate very much to our focus, 
in the form of probability-endowed connectives similar to 
implications. We must point out as well that a yearly meeting
(Uncertainty in Artificial Intelligence, these days in its
34th edition) keeps adding a steady flow of knowledge 
to the area. A general trait of most of these publications 
is that they work in a context substantially wider than ours.

Indeed, we concentrate on a much narrower focus, heavily 
influenced both by the afore-mentioned Horn logic, in its
most basic (propositional) incarnation, and by
practice-oriented data-mining frameworks, namely association
rules; in exchange for such a narrow (but still very relevant
for practice) focus, we aim at obtaining stronger theorems.

Horn formulas can be seen as conjunctions of implications 
(details below). They have been studied from the active 
learning perspective (see~\cite{AFP},~\cite{AriasBalcazarML}) 
and through their connections with closure spaces and formal
concept analysis (\cite{GW},~\cite{Wild}). They are also
closely related to inference rules for full-fledged logics 
and, in that direction,
a contribution very 
relevant to our work (as detailed below) is \cite{ParisSimmonds}, 
which, in turn,
builds on earlier work on the comparison between probabilistic 
and qualitative variants of 
inference schemes \cite{MakinsonHawthorne}. 

Given that our focus is on notions of redundancy, that
we will formalize in the form of logical entailment, we 
point out some nice, related properties of Horn formulas. 
Syntactically, it is known that a set of implications
$\B$ entails another implication $X\dto Y$ if and only if $X\dto Y$ is derivable 
from $\B$ via the Armstrong axiom schemes, namely, 
Reflexivity ($X\dto Y$ for $Y\subseteq X$), 
Augmentation (if $X\dto Y$ and $X'\dto Y'$, then $XX'\dto YY'$,
where juxtaposition denotes conjunction) 
and 
Transitivity (if $X\dto Y$ and $Y\dto Z$, then $X\dto Z$).
See the survey \cite{Wild} for details and references\footnote{An
earlier version of that survey, available at
\url{http://arxiv.org/abs/1411.6432v2}, contains 
appendices with more detailed
explanations regarding these facts than the formal journal publication.}.

Besides, out of any set of implications, it is possible to 
identify a canonical and minimum-cardinality subset 
from which the whole set can be derived
(see e.~g.~\cite{AriasBalcazarML}, \cite{GW}, and \cite{Wild}). 
In practice, its size is often amazingly small.
All of this parallels closely related work on functional 
dependencies in databases.
Within the contexts of closure spaces and data mining, 
these small sets of implications are usually called 
``bases'', whereas for
dependency 
theory they are often called ``covers''.


Both in machine learning and in data mining, one particularly
well-studied knowledge representation mechanism is given
by \emph{relaxed} implication connectives: a natural 
abstract concept which can be made concrete in various ways.
The common idea is to relax the semantics of the implication
connective so as to allow for exceptions, 
a feature actually mandatory in many applications in data analysis 
or machine learning. However, this can be done in any of
a number of ways; and each form of endowing relaxed implications 
with a precise meaning yields a different notion with, often,
very different properties. See \cite{LavracFZ99}, the survey~\cite{GH}
and the book \cite{GW};
but, again, the literature on the topic is huge: we mention
here only the references most relevant for our specific results, 
and refer for further context and additional references 
to our earlier paper~\cite{Balcazar}.

That paper, of which this one is a close follow-up, 
focuses on one of the simplest forms of relaxed
implication, endowed with its most natural semantics: the one
given by conditional probability.
Syntactically, these 
partial implications are pairs of conjunctions of positive 
propositional literals.
For sets $X$ and $Y$ of propositional variables, we write
the corresponding partial implication as $X\to Y$. 
Now, instead of the classical semantics, whereby a model 
satisfies the implication if it either fails the antecedent
or fulfills the consequent, we want to quantify exceptions;
hence, instead of individual propositional models, our semantic 
structures are, then, so-called ``transactional datasets'',
that is, multisets of propositional models. By mere counting, 
we find, on each dataset, a frequentist probability for
$X$ and $Y$ seen as conjunctions (or, equivalently, as
events): then, the meaning of the implication is, simply, 
that the conditional probability of the consequent, given
the antecedent, exceeds some fixed threshold, 
here denoted~$\g\in(0,1)$.
Very often, 
that quantity, the frequentist conditional probability,
is called \emph{confidence} of the partial implication.
We also use this name here.

This probabilistic version 
of implications has been proposed in different
research communities. For instance, \cite{Lux} 
introduced them as ``partial implications''; much later,
\cite{AIS} defined ``association rules'' (see also
\cite{AMSTV} and the surveys \cite{CegRod}, \cite{GH}): 
these are partial implications that impose the additional
condition that the consequent is a single propositional
variable, and where additional related parameters 
(prominently ``support'', defined below) are used
to assess their interest. 

Actually, confidence does not seem to be the best
choice in practice as the meaning of a partial
implication; this is discussed e.~g.~in~\cite{GH}.
However, it is clearly the most natural choice and the
obvious step to start the logical study of partial 
implications, many other preferable options being
themselves, actually, variations or sophistications 
of confidence.

Now: given a set of partial implications,
all of them true of our data at confidence threshold $\g$,
assume we wish to identify a smallish subset from which all
of them ``follow logically'' --- a task of ``redundancy suppression''
that is common in all practical applications of association rules. 
Two proposals in \cite{AgYu} and \cite{KryszPAKDD} turned out 
to be equivalent among them 
and were, in turn, as described 
in \cite{Balcazar}, equivalent to the natural notion of logical 
entailment of one partial implication by another 
(modulo minor details such as allowing 
or disallowing empty antecedents or consequents). 
This entailment means that any dataset 
in which the premise reaches confidence at least~$\g$ 
must assign confidence at least $\g$ as well to the conclusion.
The formalization is provided below, but, in essence, the
antecedent of the conclusion must include the antecedent
of the premise (so that it ``fires''), and the union of 
antecedent and consequent of the premise must include
the antecedent and consequent of the conclusion
\cite{KryszPAKDD}.

But, then, what about using more than one premise?
Entailment among partial implications is quite different from 
entailment among classical implications. 
First, Transitivity
fails: it is not difficult to see that, 
if $X\to Y$ has confidence over $\g$, and 
$Y\to Z$ as well, still most occurrences of $Y$ could be
without $X$, leaving low or even
zero confidence for $X\to Z$. Even 
if we consider $X\to Y$ and $XY\to Z$, the probabilities
multiply together and leave just $\g^2 < \g$ as provable
threshold. (Cf.~\cite{Balcazar} for further details.)

Moreover, Augmentation fails as well.
While $A \dto B$ classically entails $AC \dto BC$,
such entailment fails badly for partial implications:

\begin{exa}
Consider a dataset consisting of many transactions $AB$, 
plus exactly one transaction $AC$. Then 
$A \to B$ can have as high a confidence as desired,
by enlarging the dataset, while $AC \to BC$ has confidence zero.
\end{exa}


A tempting intuition is to generalize the
observation and jump to the statement that no nontrivial
consequence follows from two partial implications; however,
this statement is wrong. We reproduce below in 
Section~\ref{sec:uptotwopremises} a characterization from
\cite{Balcazar} of the cases of proper entailment from
two premises; for this introduction, we just present
some explicit examples, which motivate as well the
extensions developed in this paper. We employ a simple, 
famous, and relatively small dataset often used 
for teaching introductory data analysis courses. It~comes 
from data of each of the passengers of the Titanic. Among 
several existing variants of this dataset, some of them 
pretty complete, we choose a reduced variant that keeps four 
attributes, one of them (age) discretized. To describe the details 
of this dataset, we quote:

\begin{itemize}\item[]
``The titanic dataset gives the values of four categorical 
attributes for each of the 2201 people on board the Titanic 
when it struck an iceberg and sank. The attributes are social 
class (first class, second class, third class, crewmember), 
age (adult or child), sex, and whether or not the person survived.''
\end{itemize}

\rightline{(\url{http://www.cs.toronto.edu/~delve/data/titanic/desc.html})}

\smallskip
(According to that website, this variant of the
data was originally compiled by Dawson~\cite{Dawson1995}
and converted for use in the DELVE data analysis
environment by Radford Neal.)

\begin{exa}
We give first an example of the well-studied case of entailment 
of one partial implication by another.
Suppose that we analyze our dataset at a very mild confidence 
threshold of $0.54$ (with support threshold, defined below, of 1\%).
We then find the partial implication

\smallskip
Class:3rd $\to$ Age:Adult Sex:Male Survived:No

\smallskip\noindent
together with 

\smallskip
Class:3rd Sex:Male $\to$ Survived:No

\smallskip\noindent
which, in fact, can be omitted because, due to the inclusion
properties, it must have as much confidence, or more, 
as the previous one, whatever the dataset.
\end{exa}

\looseness=-1
The contributions of \cite{Balcazar} that are relevant to
the present paper are, chiefly: first, syntactic 
characterizations of one partial implication
entailing another as in this example; second, a similar fact for two 
partial implications entailing another; and, third,
the generalization of both to entailment relative to
the presence of classical implications. (Further, that
reference provides studies about minimal bases for partial 
implications.) We return to our example dataset.

\begin{exa}
\label{exa:twoprem}
In the same conditions as before, we find the partial
implication

\smallskip
Class:1st Sex:Male Survived:Yes $\to$ Age:Adult

\smallskip\noindent
that is not redundant with respect to any other
partial implication found at the same confidence
and support thresholds (we omit the details of the
process that proves this: it consists of applying
the tools in Section~\ref{sec:uptotwopremises}). 
However, it turns out to
be redundant if we consider the pair of implications,
also found at these thresholds,

\smallskip
Class:1st $\to$ Survived:Yes Age:Adult

\smallskip
Class:1st $\to$ Sex:Male Age:Adult

\smallskip\noindent
where the entailment is a far from trivial fact that
follows from a major contribution of \cite{Balcazar}
that we extend in the present paper. We grab the
opportunity to point out, though, that, in practical
datasets with real-life data, it seems to be extremely 
uncommon to find cases like this one; and that, as of
now, algorithms to efficiently decide entailment from
more than one premise are not yet available, as we discuss
at length below.
\end{exa}

In the earlier conference version of the present paper~\cite{AtsBalLICS},
we generalized nontrivially the characterization of
entailment to an arbitrary number of premises, beyond
the cases of one and two premises in \cite{Balcazar}.
The case of two premises, specifically, is behind 
Example~\ref{exa:twoprem}. The proof of the characterization
for this case in \cite{Balcazar}
is
not deep, using just basic set-theoretic constructions;
but it is long, cumbersome, and of limited intuitive value.
Attempts at generalizing it directly to more than two premises
rapidly reach unmanageable difficulties, among which the
most important one is the lack of hints at 
the right generalization of a crucial
property that we will explain below in Section~\ref{sec:high}.



Thus, in~\cite{AtsBalLICS}, 
we started the development of an alternative, quite different 
approach, that turns out to be successful in finding the 
right characterization. 
Our first ingredient is a connection with linear programming 
that is almost
identical to a technical lemma in \cite{ParisSimmonds},
which applies to all values of the confidence threshold $\g\in(0,1)$. 
Stated in our
language, the lemma asserts that $k$ partial implications entail
another one if and only if the dual of a linear program 
naturally associated to the entailment is feasible. 
We prove a number of facts related to that technical tool;
then, we use them to get our main results. These concentrate on a
study of the different situations that may appear, depending on
intervals to which $\g$ belongs:

1/ for low enough values of the confidence threshold $\gamma$, we
  show that $k$ partial implications, with $k>1$,
  never entail nontrivially another one;

2/ for high enough values of $\gamma$, we characterize exactly
  the cases in which $k$ partial implications entail another one, 
  in a manner that generalizes the approach of \cite{Balcazar};
  namely, the characterization runs purely in terms of elementary
  Boolean-algebraic conditions involving just simple set-theoretic 
  properties of the partial implications involved;
  
3/ for the intermediate values of $\gamma$, we explain how to 
  identify 
  the exact threshold, if any, at which a specific set of
  $k$ partial implications entails another one.

\looseness=1
The characterizations provide algorithms to
decide whether a given entailment holds. More concretely, under very
general conditions including the case that $\gamma$ is large, the
connection to linear programming
gives an algorithm that is polynomial in the
number of premises~$k$, but exponential in the number of attributes
$n$. Our subsequent characterization 
reverses the situation: it gives an algorithm that
is polynomial in $n$ but exponential in~$k$. 


Our main characterization also shows that the decision
problem for entailments at large~$\gamma$ is in NP, and this does not
seem to follow from the linear programming formulation by itself
(since the program is exponentially big in $n$), let alone the
definition of entailment (since the number of datasets on $n$
attributes is infinite due to the relevance of multiplicities). 
We discuss this in Section~\ref{sec:closing}.



The present, archival version of this paper includes an
additional development: relative entailment. To explain and
motivate it, let us return once more to our dataset.

\begin{exa}
We consider now partial implications in the Titanic
dataset under the slightly more demanding confidence
threshold of 0.7. We find a case that looks, on the
surface, exactly like the previous one in Example~\ref{exa:twoprem}; 
namely, we find these three partial implications:

\smallskip
Class:Crew Sex:Male Age:Adult $\to$ Survived:No

\smallskip
Class:Crew $\to$ Age:Adult Survived:No

\smallskip
Class:Crew $\to$ Sex:Male Survived:No

\smallskip\noindent
and, exactly as before, the first is entailed jointly
by the second and third.
However, in this case, if we take into account 
entailment relative to classical implications, things
change substantially. Indeed, it turns out that 100\% of the
cases obey the classical implication

\smallskip
Class:Crew $\dto$ Age:Adult

\smallskip\noindent
(that is, child labor was absent in the Titanic, 
of course, at least according to the official records). 
Taking the entailment relative to this classical implication, 
and resorting again to the tools in Section~\ref{sec:uptotwopremises},
one can find that the partial implication 

\smallskip
Class:Crew $\to$ Sex:Male Survived:No

\smallskip\noindent
suffices as premise to obtain the same conclusion

\smallskip
Class:Crew Sex:Male Age:Adult $\to$ Survived:No;

\smallskip\noindent
thus falling back into the case of one 
single partial premise, instead of two.
\end{exa}

\looseness=-1
It pays off, therefore, to consider separately the
classical implications, and to reason about entailment among
partial implications relative to classical ones.
(This idea goes back at least to~\cite{Zaki};
see further references in \cite{Balcazar}.)
Indeed this truly extends, both
in theory and in practice, the scope of applications of
the efficient case of one single partial implication as
premise. The reason is that classical 
implications can be summarized better, because they allow 
for Transitivity and Augmentation to apply in order to 
find redundancies, while these properties are unavailable
for partial implications.
Thus, we will work in the presence of some fixed, arbitrary set of
classical implications; the entailment is considered \emph{relative} 
to this set and, when this set is empty, we fall back 
into the standard case of entailment among partial implications.

In \cite{Balcazar}, this sort of relative entailment was
developed to cover up to two partial implications as 
premises, but the conference version of this paper did not
provide such a view for the case of arbitrarily many premises.
Allowing for this more general
form is not trivial, because we need to precisely identify
the exact correlate of each of our technical definitions. 
We complete
the development here, aiming at a wider-scope, self-contained 
paper that does not require to consult the conference paper.

\section{Preliminaries and notation} \label{sec:prelim}

Our expressions involve 
propositional variables, which receive Boolean
values from propositional models; we define
their semantics through datasets: simply, 
multisets of propositional models. 
However, we mostly follow a terminology
closer to the standard one in the data analysis 
community, where our propositional variables are
called attributes or, sometimes, items; likewise, 
a set of attributes (that is, a propositional model), 
seen as an element of a dataset, is often called a 
transaction.

Thus, attributes take Boolean values, true or false, and
a transaction is simply a subset of attributes, those that would be
set to true if we thought of it as a propositional model. Typically, our
set of attributes is simply~$[n] := \{1,\ldots,n\}$, for a natural
number $n>0$, so transactions are subsets of~$[n]$. Fix now such a set
of attributes.
For sets of attributes $X$, $Z$, we say that
$Z$ covers $X$ if $X \subseteq Z$. 
This term is used only when $Z$ is a transaction, and 
will be refined into various possible relationships
between transactions and partial implications.

Formally, a dataset, as a multiset of
transactions, is a mapping from the set of all transactions to
the natural numbers, namely, their multiplicities: 
transactions mapped to zero do not appear
in the dataset, while nonzero values indicate the multiplicity
with which the transaction appears in the dataset. In practical
applications, the dataset is, most often, an ordered list of 
transactions, with repeated occurrences according to the 
multiplicities, but here we prefer the more formal view where 
no ordering is unnecessarily imposed. Given dataset $\D$, 
we write $Z\in\D$ to 
indicate a nonzero multiplicity and $Z\notin\D$ for zero
multiplicity.

If $\mathcal{D}$ is a dataset and $X$ is a set
of attributes, we write $s_{\mathcal{D}}(X)$ for the so-called
\emph{support} of $X$ in $\D$: the
number of transactions in $\mathcal{D}$ that cover $X$, counted with
multiplicity; that is, the sum of all the multiplicities of transactions
$Z\in\D$ where $X\subseteq Z$. A number of key practical algorithms in Data Mining
rely on the antimonotonicity property of support:
$s_{\mathcal{D}}(X) \leq s_{\mathcal{D}}(Y)$ whenever $Y\subseteq X$.

If $X$ and $Y$ are sets of attributes, 
we write their juxtaposition $XY$ to denote 
their union $X\cup Y$. This is fully customary 
and very convenient notation in this context. 

\subsection{Classical implications and closure spaces}
\label{ss:armstrong}


Our starting point is Horn logic or, more precisely,
definite Horn clauses. A clause is a disjunction of
possibly negated propositional variables; it  
is a definite Horn clause if it contains exactly one
non-negated variable. 
All our Horn clauses are definite: 
often we omit the adjective.
A Horn formula is a conjunction
of Horn clauses. We represent Horn formulas in 
implicational form by grouping together into a single
expression $X\dto Y$ all the Horn clauses with the
set $X$ of negated attributes, each contributing their
positive attribute to $Y$. We employ liberally the
standard satisfaction and entailment symbol: 
$Z\models X\dto Y$ represents the fact that 
transaction $Z$ satisfies the implication,
either vacuously, by not covering $X$, or by covering $XY$.

We will find it useful to introduce terminology 
distinguishing these two satisfaction cases. That is, following 
standard usage 
(see e.g.~\cite{AFP}),
we say that a transaction $Z \subseteq [n]$ \emph{covers} 
an implication $X\dto Y$ if
it covers $X$: $X\subseteq Z$; and that $Z$ \emph{violates} 
the implication if 
it covers $X$ but not $Y$: $X\subseteq Z$ but
$Y\not\subseteq Z$. If $Z$ covers $X \dto Y$ without violating it, 
that is, $XY \subseteq Z$, we
say that $Z$ \emph{witnesses} $X \dto Y$.

For a set $\B$ of implications, 
$Z\models\B$ means 
$Z\models\bigwedge \{X\dto Y | (X\dto Y)\in\B\}$;
and for a dataset $\D$, $\D\models\B$ means that $Z\models\B$
for all $Z\in\D$.
We also employ the same symbol with its standard overloading:
a set of implications entails another one, in symbols
$\B\models X\dto Y$, if for every 
$Z\models\B$
we have $Z\models X\dto Y$.
As indicated in the Introduction, this happens if and 
only if $X\dto Y$ is derivable from $\B$ via the Armstrong
axiom schemes:
Reflexivity, 
Augmentation and 
Transitivity.
This gives us a clear and robust notion
of redundancy among implications, one
that can be defined equivalently both 
in semantic terms and through a syntactic calculus. 

We will need some notation about closures. 
The fact, well-known in logic and knowledge representation, 
that Horn theories are exactly those closed under 
bitwise intersection of propositional models 
leads to a strong connection with Closure Spaces,
where closure under intersection always holds
(see~the discussions in \cite{DP}~or~\cite{KR}).
A basic fact from the theory of Closure Spaces is that
closure operators are characterized by three properties:
extensivity ($X\subseteq\cls{X}$), idempotency
($\cls{\cls{X}} = \cls{X}$), and monotonicity
(if $X\subseteq Y$ then $\cls{X}\subseteq\cls{Y}$).
A set is closed if it coincides with its closure. 
Usually we speak of the {\em lattice} of closed sets
(technically it is just a semilattice in general but, 
in our case, the fact that we only employ definite
Horn clauses leads to a lattice). The bottom of
the lattice is $\overline{\emptyset}$, which 
equals $\overline{X}$ for every $X\subseteq\overline{\emptyset}$.

The connection between classical implications and
closure operators runs as follows: given~$\B$, a set
of implications, the closure $\cls{X}$ of a set $X$
is the largest set $Y$ such that $\B\models X\dto Y$
(extensivity, idempotency, and monotonicity are
easy to check);
whereas, if we are given a closure operator, 
we can axiomatize it by the set of implications
$\{ X\dto Y \mid Y \subseteq \cls{X}, X\subseteq[n] \}$
or, equivalently, any set of implications that entails
exactly this set.
Thus,
$\B\models X\dto Y$ if and only if $Y\subseteq\cls{X}$.

\begin{prop}
\label{prop:closuresdataset}
Given a dataset $\D$ and a set of implications $\B$,
with its associated closure operator mapping each itemset
$X\subseteq[n]$ to $\cls{X}$, the following are equivalent:
\begin{enumerate}
\item
$\D\models\B$,
\item
$\forall Z\in\D$, $Z = \cls{Z}$,
\item
$\forall Z\in\D$, 
$\forall X \subseteq [n]$, 
if $X\subseteq Z$ then 
$\cls{X}\subseteq Z$,
\item
$\forall X \subseteq [n]$, $s_{\D}(X) = s_{\D}(\cls{X})$.
\end{enumerate}
\end{prop}

\proof
{(1)}${}\Rightarrow{}${(2)} 
$\D \models \B$ means that $Z \models \B$, for all $Z \in D$.
But $\B\models Z\dto\cls{Z}$, so that $\cls{Z}\subseteq Z$,
which implies equality.

{(2)}${}\Rightarrow{}${(3)} 
For all $X \subseteq Z$, by monotonicity, $\cls{X} \subseteq \cls{Z} = Z$.

{(3)}${}\Rightarrow{}${(4)} 
We argue first that $\forall X \subseteq [n] (X \subseteq Z \iff \cls{X} \subseteq Z)$:
one direction is because $X\subseteq\cls{X}$ and the other by assumption.
Then, the sums of multiplicities for computing $s_{\D}(X)$  and $s_{\D}(\cls{X})$
run on the same transactions and, hence, give the same result.

{(4)}${}\Rightarrow{}${(1)} 
For every $X$, the facts that $X\subseteq\cls{X}$ and 
$s_{\D}(X)= s_{\D}(\cls{X})$ imply that $X$ and $\cls{X}$
are subsets of exactly the same transactions $Z\in\D$.
Let $X\dto Y\in\B$: then $Y\subseteq\cls{X}$. For every $Z \in D$,
if $X\subseteq Z$ then $Y\subseteq \cls{X}\subseteq Z$ so that
$Z\models X\dto Y$.
\qed

\looseness=-1
In practice, given a dataset $\D$, 
we mainly consider two options for $\B$ and for the
corresponding closure operator: namely,
$\B = \emptyset$, where
the closure operator is the identity, $\cls{X} = X$ 
for all $X$ (that is, all sets are closed); or 
$\B$ being the set of all implications that are true in $\D$
(or any basis that entails that set); in this case, the closure
of $X$ is the largest set such that $s_{\D}(X) = s_{\D}(\cls{X})$.
It is easy to prove that such a $\cls{X}$ exists and is unique.
For instance, in this case, $\cls{\emptyset}$ is exactly
the set of attributes (if any) that appear in every transaction.

Equivalently, it is also known that, 
for this second case, the closure
of itemset $X$ is the intersection of all the
transactions that contain $X$. Essentially, 
$X\subseteq\cls{X}$ implies that all
transactions contributing to the support 
of $\cls{X}$ include $X$ as well: hence,
if the support counts coincide, then they must count
exactly the same transactions
(see \cite{GW}, \cite{Zaki}, and the references therein
for precise proofs of all these statements).
Several quite good algorithms exist to find, for
a given dataset, the corresponding closed sets 
and their supports (see section~4 of~\cite{CegRod}). 

\subsection{Partial implications}

A partial or probabilistic implication consists of a pair of finite
subsets $X$ and $Y$ of attributes. We write them as $X \to Y$. 
We extend also to partial implications the terminology
introduced above:
we say that a transaction $Z \subseteq [n]$ \emph{covers} $X\to Y$ if
it covers $X$; that $Z$ \emph{violates} $X \to Y$ if 
it covers $X$ but not $Y$, and that $Z$ \emph{witnesses} $X \to Y$
if it covers $XY$.
But it is important to note that these are \emph{not} anymore
directly related to a $\models$ relationship, as the semantics
of the partial implication is different.

Specifically,
let $X \to Y$ be a partial implication with all its
attributes in $[n]$,
let $\mathcal{D}$ be a dataset on the set of attributes $[n]$, and
let $\gamma$ be a real parameter in the open interval $(0,1)$. 
We write
$\mathcal{D} \models_\gamma X \to Y$ if either
$s_{\mathcal{D}}(X) = 0$, or else 
  $s_{\mathcal{D}}(XY)/s_{\mathcal{D}}(X) \geq \gamma$. 
Equivalently, and with the advantage of not needing the
zero-test: $\mathcal{D} \models_\gamma X \to Y$ if
  $s_{\mathcal{D}}(XY) \geq \gamma s_{\mathcal{D}}(X)$,
since $s_{\mathcal{D}}(X) = 0$ implies
$s_{\mathcal{D}}(XY) = 0$ by antimonotonicity.

Thus, if we think of $\mathcal{D}$ as
  specifying the probability distribution on the set of transactions
  that assigns probabilities proportionally to their multiplicity in
  $\mathcal{D}$, then $\mathcal{D} \models_\gamma X \to Y$ if and only
  if the conditional probability of $Y$ given $X$ is at least~$\gamma$.
The real number $\gamma$ is often referred
to as the \emph{confidence} parameter.

For a partial implication $X \to Y$, its \emph{classical counterpart}
is simply and naturally $X \dto Y$.
If $X_0 \to Y_0,\ldots,X_k \to Y_k$ are partial implications, we write
\begin{equation}
X_1 \to Y_1,\ldots,X_k \to Y_k \models_\gamma X_0 \to Y_0
\label{eqn:entailmentNoB}
\end{equation}
to express that for every dataset $\mathcal{D}$ for which $\mathcal{D}
\models_\gamma X_i \to Y_i$ holds for every $i \in [k]$, it also holds
that $\mathcal{D} \models_\gamma X_0 \to Y_0$. Note that the symbol
$\models_\gamma$ is overloaded much in the same way that the symbol
$\models$ is overloaded in propositional logic.  In
case Expression~\eqref{eqn:entailmentNoB} holds, we say that the entailment holds, or
that the set $X_1 \to Y_1,\ldots,X_k \to Y_k$ entails $X_0 \to Y_0$ at
confidence threshold $\gamma$.

The extreme cases of $\g = 0$ and $\g = 1$, which are left out
of our discussion since $\g\in(0,1)$, are worth a couple of words
anyhow. Clearly $\gamma = 0$ does not provide an interesting setting:
  $s_{\mathcal{D}}(XY) \geq \gamma s_{\mathcal{D}}(X)$ 
is always true in this case, so the definition trivializes and
every $X \to Y$ is valid. On the other hand, for $\g = 1$
the semantics of $X \to Y$ is that of a classic implication,
of which we have already discussed the major properties.

As discussed informally in the Introduction, 
there may be
partial implications that do, actually, reach
confidence 1, that is, they are classical implications.
Since $\g < 1$, there is no contradiction in treating them together with
the rest; however, it has been observed (\cite{Balcazar}, \cite{Zaki})
that, in practical cases, it is worthwhile to treat them
separately, replacing them by their canonical axiomatization,
that is often very small in practice, and thus discussing the 
truly partial implications separately. 

Hence, several studies, prominently \cite{Zaki}, 
have put forward a different notion of redundancy;
namely, they give a separate role to the
full-confidence implications, often through
their associated closure operator. Along this way,
one gets a stronger notion of redundancy and, 
therefore, a possibility that smaller bases can be constructed.
We follow up this line of thought by considering 
\emph{relative entailment}; more precisely, we discuss
when entailment among partial implications holds in a
sense akin to that of Expression~\ref{eqn:entailmentNoB}, but
in the presence of a fixed set of background classical 
implications $\B$.

For this general case, we 
consider \emph{entailment relative to} $\B$ in the
following sense:
\begin{equation}
\B, X_1 \to Y_1,\ldots,X_k \to Y_k \models_\gamma X_0 \to Y_0.
\label{eqn:entailment}
\end{equation}
That is, in all datasets that satisfy (classically, of course)
the classical
implications $\B$ and that give confidence at least $\g$
to the $k$ partial implication premises, the partial
implication in the conclusion also must reach confidence 
at least $\g$.
Equivalently, at the time of discussing entailment
as in Expression~\ref{eqn:entailment}, we restrict
our discussion to datasets $\D$ that satisfy $\B$.
The most interesting case is, of course, when $\B$
is (equivalent to) the set of all the classic 
implications that hold in a given dataset.
On the other hand, 
for the particular case of $\B = \emptyset$, already
mentioned, of course we fall back into 
Expression~\ref{eqn:entailmentNoB} at
its face value. 

\looseness=-1
If $\Sigma$ is a set of partial
implications for which $\Sigma \models_\gamma X_0 \to Y_0$ holds, but
$\Gamma \models_\gamma X_0 \to Y_0$ does not hold for any proper
subset $\Gamma \subset \Sigma$, then we say that the entailment holds
properly, with the corresponding variant for the relative case.
Note that entailments without premises vacuously hold
properly when they hold. Of course, an improper entailment
can be transformed into a proper one by simply omitting the
unnecessary premises: 

\noindent%
\begin{minipage}{\textwidth}%
\begin{prop}
\label{prop:properent}
The following are equivalent:
\begin{enumerate} 
\item $\B, X_1 \to Y_1,\ldots,X_k \to Y_k \models_\gamma X_0 \to Y_0$;
\item there is a set $L \subseteq [k]$ such that 
$\B,\{ X_i \to Y_i : i \in L \} \models_\gamma X_0 \to Y_0$ holds properly.
\end{enumerate}
\end{prop}
\end{minipage}

\proof
That {(2)}~implies {(1)}~is clear from the definition.
It is also easy to see that {(1)}~implies {(2)}: 
  the family of all sets
  $L\subseteq [k]$ for which the entailment 
  $\B, \{ X_i \rightarrow Y_i : i \in L \} \models_\gamma X_0 \rightarrow Y_0$ 
  holds is non-empty,
  as {(1)}~says that~$[k]$ belongs to it. Since it is finite, it
  has minimal elements, and it suffices to pick one of them for $L$.
\qed

\subsection{Linear programs}

A linear program (LP) is the following optimization problem: $\min \{
c^\mathrm{T} x : Ax \geq b,\, x \geq 0 \}$, where $x$ is a vector of
$n$ real variables, $b$ and $c$ are vectors in $\mathbb{R}^m$ and
$\mathbb{R}^n$, respectively, and $A$ is a matrix in
$\mathbb{R}^{m\times n}$. The program is feasible if there exists an
$x \in \mathbb{R}^n$ such that $Ax \geq b$ and $x \geq 0$. The program
is unbounded if there exist feasible solutions with arbitrarily small
values of the objective function $c^\mathrm{T} x$. If the goal were
$\max$ instead of $\min$, unboundedness would refer to arbitrarily
large values of the objective function. The dual LP is $\max\{
b^\mathrm{T} y : A^{\mathrm{T}} y \leq c,\, y \geq 0 \}$, where $y$ is
a vector of $m$ real variables. Both LPs together are called a
primal-dual pair. The duality theorem of linear programming states
that exactly one of the following holds: either both primal and dual
are infeasible, or one is unbounded and the other is infeasible, or
both are feasible and have optimal points with the same optimal value. 
(See \cite{Karloff}, Corollary 25 and Theorem 23.)


\section{Previous work and some related facts} 

We review here connected existing work. We describe first
the results from \cite{Balcazar} on entailments among
partial implications with one or two premises. The study there starts
with a detailed comparison of entailment as defined in
Section~\ref{sec:prelim} with the notions of redundancy among partial
implications previously considered in the literature. Also,
that reference works permanently under the assumption that a
set of classical implications, with their corresponding closure
space, is present. Here we consider first entailment as defined in
Section~\ref{sec:prelim}; for simplicity, 
we just review, for the time being, the particular
case where no background implications apply, 
so that the associated closure
operator is the mere identity: every set is closed.
The actual statement relative to background classical
implications is postponed to a later section 
(Theorem~\ref{th:case1withB}).
Then, we develop a variant of a result in \cite{ParisSimmonds},
adapted to our context and notation, on which our main
results are based, plus additional properties related
to that variant.

\subsection{Up to two premises} \label{sec:uptotwopremises}

It can be easily checked that 
the case of zero premises, i.e.~tautological
partial implications, trivializes to the classical case:
$\models_\gamma X_0 \rightarrow Y_0$ if and only if $Y_0 \subseteq
X_0$, at any positive confidence threshold $\gamma$. The first interesting
case is thus the entailment from one partial implication $X_1 \to Y_1$
to another $X_0 \to Y_0$. If $X_0 \to Y_0$ is tautological by itself,
there is nothing else to say. Otherwise, entailment is still
characterized by a simple Boolean-algebraic condition on the sets
$X_0$, $Y_0$, $X_1$, and $Y_1$ as stated in the following theorem:

\begin{thmC}[\cite{Balcazar}] \label{th:case1}
Let $\gamma$ be a confidence parameter in $(0,1)$ and let $X_0 \to Y_0$
and $X_1 \to Y_1$ be two partial implications. Then the following are
equivalent:
\begin{enumerate} 
\item $X_1 \to Y_1 \models_\gamma X_0 \to Y_0$,
\item either $Y_0 \subseteq X_0$, or $X_1 \subseteq X_0$ and $X_0Y_0 \subseteq X_1Y_1$.
\end{enumerate}
\end{thmC}

\noindent 
Note that the second statement is independent of $\g$. 
This shows that entailment at confidence $\gamma$ below $1$
differs from classical entailment, as we have already
pointed out earlier.

The case of two partial implications entailing a third was also solved
in \cite{Balcazar}. The starting point for that study was a specific
example of a non-trivial entailment:
\begin{equation}
A\to BC, \, A\to BD \models_{1/2} ACD\to B. \label{eqn:exampleoftwo}
\end{equation}
Indeed, this entailment holds true at any $\gamma$ in the interval
$[1/2,1)$. This is often found counterintuitive. A common intuition 
is that combining two partial implications that only guarantee the
threshold $\g < 1$ would lead to arithmetic operations leading to values
unavoidably below~$\g$, as it happens in our earlier discussions of
Augmentation and Transitivity. However, this intuition is incorrect, 
as Expression~\eqref{eqn:exampleoftwo} shows. The good news is that 
a similar statement, when appropriately generalized, covers all the 
cases of entailment from two partial implication premises.  
We omit the proof of Expression~\eqref{eqn:exampleoftwo} as it follows
from the next theorem, which will be generalized below in our main result.

\begin{thmC}[\cite{Balcazar}] \label{th:case2}
Let $\gamma$ be a confidence parameter in $(0,1)$ and let $X_0 \to Y_0$,
$X_1 \to Y_1$ and $X_2 \to Y_2$ be three partial implications.
If $\gamma \geq 1/2$, then the following are equivalent:
\begin{enumerate} 
\item $X_1 \to Y_1,\, X_2 \to Y_2\, \models_{\gamma} X_0 \to Y_0$,
\item either $Y_0 \subseteq X_0$, or $X_i \subseteq X_0$ and $X_0Y_0
  \subseteq X_iY_i$ for some $i \in \{1,2\}$, or all seven
  inclusions below hold simultaneously:
\begin{enumerate} 
\item
$X_1 \subseteq X_2Y_2$ and $X_2 \subseteq X_1Y_1$,
\item
$X_1 \subseteq X_0$ and $X_2 \subseteq X_0$,
\item
$X_0 \subseteq X_1X_2Y_1Y_2$,
\item
$Y_0 \subseteq X_0Y_1$ and
$Y_0 \subseteq X_0Y_2$.
\end{enumerate}
\end{enumerate}
\end{thmC}

\noindent 
Indeed, the characterization is even tighter than what this statement
suggests: whenever $\gamma < 1/2$, it can be shown that entailment
from two premises holds only if it holds from one or zero premises.
This was also proved in \cite{Balcazar}, thus fully covering all cases
of entailment with two premises and all confidence parameters
$\gamma$. Clearly, all conditions stated in the theorem are
easy to check by an algorithm running in time $O(n)$, where $n$ is the
number of attributes, if the sets are given as bit vectors, say.

As already indicated, the original versions of these theorems in
\cite{Balcazar} are somewhat more general: they are stated for
\emph{relative entailment}, that is, assuming a possibly nontrivial 
closure space; but they do have our statements so far as 
particular cases ($\B = \emptyset$ as discussed previously).

The proof of Theorem~\ref{th:case2} in \cite{Balcazar} is rather long
and somewhat involved, although it uses only elementary Boolean-algebraic 
manipulation. For instance, several different
counterexamples to the entailment are built \emph{ad hoc} depending on which
of the seven set-inclusion conditions fail. Its intuition-building
value is, actually, pretty limited, and a generalization to the case of
more than two premises remained elusive for quite some time. 
A somewhat subtle point
about Theorem~\ref{th:case2} is that the seven inclusion conditions
alone do not characterize \emph{proper} entailment 
(even if $\gamma \geq 1/2$, that is): 
they are only necessary conditions for that. But when
these necessary conditions for proper entailment are disjuncted with
the necessary and sufficient conditions for improper entailment, what
results is an \emph{if and only if} characterization of
entailment. That is why the theorem is stated as it is, with the two
``escape'' clauses at the beginning of part {(2)}. Our main result will
have a similar flavour, but with fewer cases to consider.

Before we move on to larger numbers of premises, one more comment is
in order. Among the seven set-inclusion conditions in the statement of
Theorem~\ref{th:case2}, those in the first item $X_1 \subseteq X_2Y_2$
and $X_2 \subseteq X_1Y_1$ are by far the least intuitive. We present
a little bit of additional information about them.

\jlb{The coming propositions may fit better in the "nicety" section,
or maybe should be removed altogether. Or maybe leave the statements
but omit the proofs.}

\begin{prop} Assume $Y_0\not\subseteq X_0$. 
Then, property $X_2 \subseteq X_1Y_1$ is equivalent to:
$X_1 \to Y_1 \models_{\gamma} X_1X_2 \to Y_1$.
\end{prop} 

\proof 
By Theorem~\ref{th:case1},
the entailment is equivalent to the conjunction of 
$X_1\subseteq X_1X_2$, which holds, and 
$X_1X_2Y_1 \subseteq X_1Y_1$, which is 
equivalent to $X_2 \subseteq X_1Y_1$.  
\qed

\begin{prop} The properties $X_1 \subseteq X_2Y_2$ 
and $X_2 \subseteq X_1Y_1$, jointly, are equivalent to:
$(X_1 \dto Y_1) \land (X_2 \dto Y_2) \models (X_1 \dto X_1Y_1X_2Y_2) \land (X_2 \dto X_1Y_1X_2Y_2)$.
\end{prop} 

\proof
Let $Z \models (X_1 \dto Y_1) \land (X_2 \dto Y_2)$ and 
assume $X_1\subseteq Z$: then we must have as well
$Y_1\subseteq Z$, hence $X_2\subseteq Z$, hence
$Y_2\subseteq Z$; the other implication is argued
symmetrically. Conversely, if $X_1 \not\subseteq X_2Y_2$
then $Z = X_2Y_2$ satisfies both $X_1 \dto Y_1$ and 
$X_2 \dto Y_2$ (in different ways) but violates 
$X_2 \dto X_1Y_1X_2Y_2$, and symmetrically for the
other possibility.
\qed 

Discovering
the right generalization of these properties 
turned out to be the key to getting
our results. This is discussed in Section~\ref{sec:high}.
Before that, however, we need to discuss a
characterization of entailment in terms of linear programming duality.
Interestingly, LP will end up disappearing altogether from the
statement that generalizes Theorem~\ref{th:case2}; its use will merely
be a (useful) technical detour.

\subsection{Entailment in terms of linear programming} 

The goal in this section is to
discuss 
valid entailments 
as in Expression~\eqref{eqn:entailment},
where each $X_i \rightarrow Y_i$ is a partial implication on the set
of attributes $[n]$,
in terms of linear programming and duality.
The characterization can be seen as a variant, 
stated in the standard form of linear programming and
tailored to our setting, of Proposition~4
in \cite{ParisSimmonds}, where it applies to 
deduction rules of probabilistic consequence
relations in general propositional logics.
The linear programming formulation makes it
easy to check a number of simple properties of
the solutions of the dual linear program at play,
which are necessary for our application (Lemma~\ref{prop:chaos}).

Before we state the characterization, we want to
give some intuition for what to expect. 
Our versions of the main theorems below will allow for the presence
of a background set of classical implications and their
corresponding closure space. 
However, for the sake of building intuition, 
we describe first just the case where the background
set of classical implications is empty and the closure operator
is the identity, as we did 
in the previous subsection.
We leave 
to Section~\ref{sec:withbackground} the discussion of the
general versions.


For each partial implication $X \rightarrow Y$ and each transaction
$Z$, we define a weight $w_Z(X \rightarrow Y)$ that, intuitively,
measures the extent to which $Z$ witnesses $X \rightarrow Y$. 
Moreover, since we are aiming to capture confidence threshold $\gamma$,
we assign the weight proportionally:
$$
\begin{array}{lll} 
  w_Z(X \to Y) = 1-\g & 
                      \text{ if } Z \text{ witnesses } X \to Y, \\
  w_Z(X \to Y) = -\g  & 
                      \text{ if } Z \text{ violates } X \to Y, \\
  w_Z(X \to Y) = 0    & 
                      \text{ if } Z \text{ does not cover } X \to Y. 
\end{array}
$$

With these weights in hand, we give a quantitative interpretation to
the entailment in Expression~\eqref{eqn:entailmentNoB}.

First note that the weights are defined in such a way that, as long as
$\gamma > 0$, a transaction $Z$ satisfies the implication $X \to Y$
interpreted classically if and only if $w_Z(X \to Y) \geq 0$. With
this in mind, the entailment in Expression~\eqref{eqn:entailmentNoB}, interpreted
classically, would read as follows: for all $Z$, whenever all weights
on the left are non-negative, the weight on the right is also
non-negative. Of course, a sufficient condition for this to hold would
be that the weights on the right are bounded below by some
non-negative linear combination of the weights on the left, uniformly
over $Z$. What the characterization below says is that this sufficient
condition for classical entailment is indeed necessary and sufficient
for entailment at confidence threshold $\gamma$, if the weights are chosen
proportionally to $\gamma$ as above. Formally:

\begin{thm}
\label{th:mainLPwithoutB}
Let $\gamma$ be a confidence parameter in $(0,1)$, and let 
$X_0 \to Y_0, \ldots,X_k \to Y_k$ be a set of partial implications.
The following are equivalent:
\begin{enumerate} 
\item $X_1 \to Y_1,\ldots, X_k \to Y_k \models_\gamma X_0\to Y_0$
\item There is a vector $\l = (\l_1,\ldots,\l_k)$ of 
real non-negative
components such that for all $Z \subseteq [n]$
\begin{equation}
\sum_{i=1}^k \lambda_i \cdot w_Z(X_i \rightarrow Y_i) \leq w_Z(X_0
\rightarrow Y_0). 
\label{eqn:inequalitieswithoutB}
\end{equation}
\end{enumerate}
\end{thm}

As already discussed, we wish a characterization able to 
encompass the case where the premises are made up, jointly,
by a set of partial implications and an additional set of 
classical implications. Theorem~\ref{th:mainLPwithoutB} will
follow as a corollary, for the particular case where $\B = \emptyset$. 
Any reader
interested in the less general but slightly easier development
corresponding to not treating classical implications separately
may check the corresponding proofs out in \cite{AtsBalLICS}.

\subsection{Characterization in the presence of classic implications}
\label{sec:withbackground}

Thus, we move on to explain what happens in the general case.
Now our premises come in two parts: a (possibly empty) set of
classic implications $\B$ plus $k$ partial implications.
The scheme is exactly as before, but now we want to ``erase from
the picture" sets that are not closed under $\B$, as we will 
want to characterize an entailment that imposes $\B$ as premises.
Due to this, the new version of the weights is:
\begin{defi}
$$ w_Z(X\rightarrow Y) = \left\{ 
\begin{array}{ll}
		1-\gamma &$ if $Z = \overline{Z}$ and $Z$ witnesses $X \to Y,\\ 
		-\gamma &$ if $Z = \overline{Z}$ and $Z$ violates $X \to Y,\\
		0 &$ if $Z\neq \overline{Z}$ or $Z$ does not cover $X \to Y.
\end{array} 
\right.$$
\end{defi}

\noindent
where the closure operator is the one associated to $\B$.
That is, it is exactly as before, but only for closed sets.
Nonclosures are to be ignored, as they do not obey the
implications, and the way of making them irrelevant is,
of course, by setting their weight to zero. 

We will remain in the general case for the rest of the paper.
Hence, the previously given definitions of the weights are to
be fully replaced by the new version. The more 
general version of the characterization is now:

\begin{thm}
\label{th:mainLP}
Let $\gamma$ be a confidence parameter in $(0,1)$, let 
$X_0 \to Y_0, \ldots,X_k \to Y_k$ be a set of partial implications and 
let $\B$ be a set of implications. The following are equivalent:
\begin{enumerate} 
\item $\B, X_1 \to Y_1,\ldots, X_k \to Y_k \models_\gamma X_0\to
  Y_0$
\item There is a vector $\l = (\l_1,\ldots,\l_k)$ of 
real non-negative
components such that for all $Z \subseteq [n]$
\begin{equation}
\sum_{i=1}^k \lambda_i \cdot w_Z(X_i \rightarrow Y_i) \leq w_Z(X_0\rightarrow Y_0). 
\label{eqn:inequalities}
\end{equation}
\end{enumerate}
\end{thm}




Towards the proof of Theorem~\ref{th:mainLP}, let us state a useful
lemma. This gives an alternative understanding of the weights 
$w_Z(X\rightarrow Y)$ than the one given above. 

\begin{lem}
\label{lm:satisfies}
Let $\gamma$ be a confidence parameter in 
$(0,1)$, let $X \rightarrow Y$ 
be a partial implication, let $\B$ be a set of implications, let $\D$ be a 
dataset that satisfies $\B$,  
and for each $Z \subseteq [n]$ let $x_Z$ be the multiplicity
of $Z$ in $\D$, that is, the number of times that $Z$ appears
as a complete transaction in $\D$. Then 
\begin{center}
$\D \models _\gamma X\rightarrow Y \Leftrightarrow \sum_{Z \subseteq \left[n\right]} w_Z(X\rightarrow Y)\cdot x_Z \geq 0$.
\end{center}
\end{lem}

\proof
We introduce some notation.
We distribute all the transactions $Z\in\D$ into
three sets, according to whether they cover, witness, or
violate the partial implication.
Let $U = \{ Z \in \D \mid X \subseteq Z \}$, 
    $V = \{ Z \in \D \mid X \subseteq Z, Y \nsubseteq Z \}$, 
and $W = \{ Z \in \D \mid XY \subseteq Z\}$. 
Note that $U = V\cup W$ and $V\cap W = \emptyset$.
Note also that
$s_{\mathcal{D}}(X) = \sum_{Z \in U} x_Z$ and, likewise, 
$s_{\mathcal{D}}(XY) = \sum_{Z \in W} x_Z$.
Hence, the fact that $\D \models _\gamma X\rightarrow Y$ means that  
$\sum_{Z \in W} x_Z \geq \gamma\sum_{Z \in U} x_Z$, that is,
$$
\sum_{Z \in W}{x_Z} - \gamma \left( \sum_{Z \in W}{x_Z}+ \sum_{Z \in V}{x_Z}\right) \geq 0.
$$

Reordering the terms, the left-hand side equals
\begin{eqnarray*}
(1 - \g) \cdot \sum_{Z \in W}{x_Z} - \g \cdot \sum_{Z \in V}{x_Z}  & = &
\sum_{Z \in W}{(1 - \g) \cdot x_Z}   + \sum_{Z \in V}{(-\g) \cdot x_Z} \\ 
&  = & \sum_{Z \in W}{w_Z(X \to Y) \cdot x_Z} + \sum_{Z \in V}{w_Z(X\to Y) \cdot x_Z}\\
& = & \sum_{Z \in U} w_Z(X\rightarrow Y)\cdot x_Z 
\end{eqnarray*}
Now, the product $w_Z(X \to Y)\cdot x_Z$ is zero, hence irrelevant for
any sum, in two cases: when $Z\in\D$ but $Z\notin U$, because it does 
not cover $X$ and $w_Z(X\to Y) = 0$, and when $Z\notin\D$, 
hence $x_Z = 0$. (There is, actually,
a potential third case of $w_Z(X\to Y) = 0$, namely when $Z \neq\overline{Z}$ 
according to $\B$, but no transaction falls in this case because
$\D\models \B$, thus this case is covered by $Z\notin\D$.) All in all,
$\sum_{Z \subseteq \left[n\right]} w_Z(X\rightarrow Y)\cdot x_Z 
=
\sum_{Z \in U} w_Z(X\rightarrow Y)\cdot x_Z$.

Therefore, we have proved that $\D \models _\gamma X\rightarrow Y$
if and only if
$\sum_{Z \subseteq \left[n\right]} w_Z(X\rightarrow Y)\cdot x_Z \geq 0$
as desired. 
%
%
%
\qed

This lemma is parallel to the first part of the
proof of Proposition~4 in \cite{ParisSimmonds}.
With this lemma in hand we can prove Theorem~\ref{th:mainLP}.
We resort to duality here, while the version in 
\cite{ParisSimmonds} uses instead the closely related
Farkas' Lemma.

\proof[Proof of Theorem~\ref{th:mainLP}]

  The statement of Lemma~\ref{lm:satisfies} leads to a natural linear
  program: for every $Z$, let $x_Z$ be a non-negative real variable;
  impose on these variables the inequalities from
  Lemma~\ref{lm:satisfies} for $X_1 \to Y_1$ through $X_k\to Y_k$, and
  check whether the corresponding inequality for $X_0 \to Y_0$ can be
  falsified by minimizing its left-hand side:

\begin{equation*}
\begin{array}{rll}
\textbf{P}: \quad \text{min} & \sum_{Z\subseteq \left[ n\right]}{w_Z(X_0\rightarrow Y_0)}\cdot x_Z         & \\
                 \text{s.t.} & \sum_{Z\subseteq \left[ n\right]}{w_Z(X_i\rightarrow Y_i)}\cdot x_Z \geq 0, & \forall i \in \left[k\right],\\
                      &                                                                 x_Z \geq 0, & \forall Z.
\end{array}
\end{equation*}

Observe that \textbf{P} is always feasible: the all-zero vector is
always a feasible solution.
The dual \textbf{D} of \textbf{P} has one non-negative variable $y_i$
for every $i \in \left[k\right]$, and one inequality constraint 
for each non-negative variable $x_Z$. Since the objective function 
of \textbf{D} would just be the trivial constant function 0, we write 
directly as a feasibility problem:
 
\begin{equation*}
\begin{array}{rl}
\textbf{D}: \quad \sum_{i \in \left[ k\right]} w_Z(X_i\rightarrow Y_i)\cdot y_i \leq w_Z(X_0\rightarrow Y_0), & \forall Z \hfill\\ 
                                                                                                  y_i \geq 0, & \forall i \in [k]. 
\end{array}
\end{equation*}

This is the characterization statement that we are trying to prove, 
replacing $y_i$ with~$\lambda_i$. Thus, the theorem will be proved if we show that the following are equivalent:
\begin{enumerate}
\item $\B, X_1\rightarrow Y_1,...,X_k\rightarrow Y_k \models_\gamma X_0\rightarrow Y_0$,
\item the primal \textbf{P} is (feasible and) bounded below,
\item the dual \textbf{D} is feasible.
\end{enumerate} 


(1)${}\Rightarrow{}$(2) 
Let us prove the contrapositive. 
Assume that \textbf{P} is unbounded below. Let $x_Z$  be a feasible 
solution with 
$\sum_{Z \subseteq \left[ n \right]} {w_Z(X_0\rightarrow Y_0)}\cdot x_Z < 0$. 
We may assume that $x_Z$ has rational components with a positive 
common denominator $N$, while preserving feasibility and a negative 
value for the objective function. Then $N\cdot x_Z$ is still a feasible 
solution and its components are natural numbers. 
Also, for $Z$ such that $w_Z(X_0\rightarrow Y_0) = 0$ the value
of $x_Z$ is irrelevant, and we fix it to $x_Z = 0$ as well;
note that this includes all cases of $Z\neq\cls{Z}$.

Let $\D$ be a transactions multiset consisting of 
$N\cdot x_Z$ copies of $Z$ for every $Z \subseteq \left[n\right]$. 
As just indicated, for $Z\neq\cls{Z}$ we add zero transactions
so that $Z = \cls{Z}$ whenever $Z\in\D$, that is, $\D\models\B$
by Proposition~\ref{prop:closuresdataset}.

By feasibility we have $\sum_{Z \subseteq \left[n \right]}{w_Z(X_i\rightarrow Y_i)}\cdot N\cdot x_Z \geq 0$ 
and therefore $\D \models_\gamma X_i \rightarrow Y_i$ for every $i \in \left[k\right]$ by Lemma~\ref{lm:satisfies}. 
On the other hand $\sum_{Z \subseteq \left[n \right]}{w_Z(X_0\rightarrow Y_0)}\cdot N\cdot x_Z < 0$ from which it 
follows that $\D \not\models_\gamma X_0 \rightarrow Y_0$, again by Lemma~\ref{lm:satisfies}.  

(2)${}\Rightarrow{}$(3) Direct consequence of the duality theorem.

(3)${}\Rightarrow{}$(1) Assume \textbf{D} is feasible and let $y$ be a feasible solution. Let $\D$ be a transactions multiset such that $\D \models \B$ and $\D\models_\gamma X_i \to Y_i$, for every $i \in [k]$. For every $Z \subseteq [n]$, let $x_Z$ be the multiplicity of $Z$ in $\D$. Since $y$ is a feasible solution and $x_Z$ is non-negative, we have:

\begin{eqnarray*}
\sum_{Z\subseteq \left[n\right]}w_Z(X_0\rightarrow Y_0)\cdot x_Z & \geq & \sum_{Z\subseteq \left[n\right] }\left( \sum_{i\in \left[k\right]}w_Z(X_i\rightarrow Y_i)\cdot y_i \right) \cdot x_Z \\
                                                                 &   =  & \sum_{i\in \left[k\right]}{y_i} \cdot \left(\sum_{Z\subseteq \left[n\right]}w_Z(X_i\rightarrow Y_i)\cdot x_Z\right)
\end{eqnarray*}

This is not negative since $y_i$ is not negative and also $\sum_{Z\subseteq \left[n\right]}w_Z(X_i\rightarrow Y_i)\cdot x_Z$ is not negative by the assumption on $\D$ and Lemma~\ref{lm:satisfies}. This proves that $\sum_{Z\subseteq \left[n\right]}w_Z(X_0\rightarrow Y_0)\cdot x_Z \geq 0$, from which $\D\models_\gamma X_0 \rightarrow Y_0$ once more by Lemma~\ref{lm:satisfies}.  
%
\qed

The sort of argumentations deployed so far will be pervasive in what
follows. However, instead of sums along $Z\subseteq[n]$, as in
the primal form, we will mostly find sums along $i\in[k]$ as in the
dual formulation. Again, it will be helpful to factor out the most common
algebraic manipulations into a technical lemma. We employ now 
the following notational variant:

\begin{defi}
Given $k$ partial implications $X_i \to Y_i$ for $i\in [k]$:
\begin{eqnarray*}
V_Z &=& \{ i \in [k] : Z \text{ violates } X_i \to Y_i \}, \\
W_Z &=& \{ i \in [k] : Z \text{ witnesses } X_i \to Y_i \}, \\
U_Z &=& \{ i \in [k] : Z \text{ covers } X_i \to Y_i \} \text{ so } U_Z = V_Z \cup W_Z. \\
\end{eqnarray*}
\end{defi}

\begin{lem}
\label{lm:mainLPBase}
Let $\gamma$ be a confidence parameter in (0,1), let $X_1 \to Y_1, \ldots, X_k \to Y_k$ be a set of partial implications, let $\B$ be a set of implications, let $\lambda = \lambda_1, \ldots, \lambda_k$
be a vector of non-negative reals, let $Z \subseteq [n]$ be such that $\overline{Z}=Z$, and let $\Gamma = \sum_{i \in [k]} \lambda_i \cdot w_Z(X_i \rightarrow Y_i)$,
that is, the left-hand side of Inequality~\ref{eqn:inequalities}. Then
\begin{enumerate} 
\item 
$\Gamma = (1-\gamma) \cdot \sum_{i \in W_Z}{\lambda_i} - \gamma \cdot\sum_{i \in V_Z}{\lambda_i}$ (whence $\Gamma\geq - \gamma \cdot\sum_{i \in V_Z}{\lambda_i}$).
\item  
$\Gamma = \sum_{i \in W_Z}{\lambda_i} - \gamma \cdot\sum_{i \in U_Z}{\lambda_i}$.
\item 
$\Gamma \geq \lambda_j - \gamma \cdot\sum_{i \in U_Z}{\lambda_i}$ for all $j\in W_Z$.
\item 
$\Gamma \leq \sum_{i \in U_Z, i\neq j}{\lambda_i} - \gamma \cdot\sum_{i \in U_Z}{\lambda_i}$ for all $j \in V_Z$
\end{enumerate}
\end{lem}

\begin{proof}
{(1)} First, we split the sum according to $U_Z$:
$$ \sum_{i=1}^k \lambda_i \cdot w_Z(X_i \rightarrow Y_i) = 
\sum_{i \in U_Z} \lambda_i \cdot w_Z(X_i \to Y_i) + \sum_{i \notin U_Z} \lambda_i \cdot w_Z(X_i \to Y_i) $$

If $i \notin U_Z$ then $w_Z(X_i \to Y_i) = 0$, so that $\sum_{i \notin U_Z} \lambda_i \cdot w_Z(X_i \to Y_i) = 0$. Therefore,

\begin{eqnarray*}
\sum_{i=1}^k \lambda_i \cdot w_Z(X_i \rightarrow Y_i) &=& \sum_{i \in U_Z} \lambda_i \cdot w_Z(X_i \to Y_i) \\
&=& \sum_{i \in W_Z} \lambda_i \cdot w_Z(X_i \to Y_i) + \sum_{i \in V_Z} \lambda_i \cdot w_Z(X_i \to Y_i) \\
&=& (1-\gamma) \cdot \sum_{i \in W_Z}{\lambda_i} - \gamma \cdot\sum_{i \in V_Z}{\lambda_i}.
\end{eqnarray*}

Since $\gamma \in (0,1)$ and $\lambda_i \geq 0$, $(1-\gamma) \cdot \sum_{i \in W_Z}{\lambda_i} - \gamma \cdot\sum_{i \in V_Z}{\lambda_i}\geq - \gamma \cdot\sum_{i \in V_Z}{\lambda_i} $.

\noindent
{(2)} By {(1)}, 
$\sum_{i=1}^k \lambda_i \cdot w_Z(X_i \rightarrow Y_i) = (1-\gamma) \cdot \sum_{i \in W_Z}{\lambda_i} - \gamma \cdot\sum_{i \in V_Z}{\lambda_i}$.
Then:
\begin{eqnarray*}
\sum_{i=1}^k \lambda_i \cdot w_Z(X_i \rightarrow Y_i) &=& \sum_{i \in W_Z}{\lambda_i} -\left( \gamma\cdot\sum_{i \in W_Z}{\lambda_i} \right)- \left(\gamma \cdot\sum_{i \in V_Z}{\lambda_i} \right)\\
&=& \sum_{i \in W_Z}{\lambda_i} -\gamma\cdot \left( \sum_{i \in W_Z}{\lambda_i}+ \sum_{i \in V_Z}{\lambda_i} \right) \\
&=& \sum_{i \in W_Z}{\lambda_i} -\gamma\cdot \sum_{i \in U_Z}{\lambda_i}. \\
\end{eqnarray*}

\noindent
{(3)} Since $j \in W_Z$ and $\lambda_j \geq 0$, 
and all weights are at least $-\g$,
the right-hand side of the inequality is at least : 
\begin{eqnarray*}
\sum_{i=1}^k \lambda_i \cdot w_Z(X_i \rightarrow Y_i) &=& \sum_{i \in U_Z} \lambda_i \cdot w_Z(X_i \to Y_i) \\
& \geq & (1-\gamma)\cdot \lambda_j + \left( (-\gamma) \cdot \sum_{i \in U_Z - \{j\}} \lambda_i \right) \\
&=& \lambda_j -(\gamma\cdot \lambda_j) - \left( \gamma \cdot \sum_{i \in U_Z - \{j\}} \lambda_i \right) = \lambda_j -\gamma \sum_{i \in U_Z } \lambda_i. \\
\end{eqnarray*}

\noindent
{(4)} Since $j \in V_Z$ and $\lambda_j \geq 0$, 
and all weights are at most $1-\g$,
we have: 
\begin{align*}
\sum_{i=1}^k \lambda_i \cdot w_Z(X_i \rightarrow Y_i) &= \sum_{i \in U_Z} \lambda_i \cdot w_Z(X_i \to Y_i) \\
 &\leq (1 -\gamma)\sum_{i \in U_Z- \left\{j \right\}} \lambda_i - \gamma \lambda_j \\
  & =   \sum_{i \in U_Z - \left\{j \right\}}\lambda_i - \left( \gamma \cdot \sum_{i \in U_Z- \left\{j \right\}}\lambda_i \right)- (\gamma \cdot \lambda_j) \\
  & =   \sum_{i \in U_Z - \left\{j \right\}}\lambda_i -\gamma\sum_{i \in U_Z}\lambda_i.
         \tag*{\qedhere}
\end{align*}
\end{proof}

\subsection{Properties of the LP characterization}

Whenever an entailment 
holds properly,
the characterization in Theorem~\ref{th:mainLP} gives a good deal of
information about the inclusion relationships that the sets satisfy,
and about the values that the $\l_i$ can take.  In this section we
discuss this.

\begin{lem}
\label{prop:chaos}
Let $\gamma$ be a confidence parameter in $(0,1)$, 
let $X_0 \to Y_0,\ldots,X_k \to Y_k$ be a set of partial implications 
with $k \geq 1$, and let $\B$ be a set of implications.
Assume that the entailment $\B, X_1 \to Y_1, \ldots, X_k \to
Y_k \models_\gamma X_0 \to Y_0$ holds properly.  In particular, $Y_0
\not\subseteq \overline{X_0}$. Let $\l = (\l_1,\ldots,\l_k)$ denote any vector as
promised to exist by Theorem~\ref{th:mainLP} for this entailment. 
The following hold:
\begin{enumerate} 
\item
$\l_i > 0$ for every $i \in [k]$. 
\item
$X_0Y_0 \subseteq \overline{X_1Y_1 \cdots X_kY_k}$.
\item
$\sum_{i \in [k]} \lambda_i \leq 1$.
\item
$X_i \subseteq \overline{X_0}$ for every $i \in [k]$.
\item
$X_iY_i \not\subseteq \overline{X_0}$ for every $i \in [k]$. 
\item
$\sum_{i \in [k]} \lambda_i = 1$.
\item $Y_0 \subseteq \overline{X_0Y_i}$ for every $i \in [k]$.
\end{enumerate}
\end{lem}

\jlb{enumeration to be formatted so as to match the statement}

\proof
{(1)} If any $\lambda_i$ is zero then, by Theorem~\ref{th:mainLP}, 
applied to the smaller set of partial implications whose 
corresponding coefficients are nonzero,
the entailment would not be a proper entailment.

{(2)} Let $Z= \overline{X_1Y_1...X_kY_k}$ then $\overline{Z} = \overline{\overline{X_1Y_1...X_kY_k}} = \overline{X_1Y_1...X_kY_k} = Z$ by idempotency, 
so we have $ \forall i \in [k], w_Z(X_i\rightarrow Y_i)= 1 - \gamma$. 
If $X_0Y_0 \nsubseteq Z$, then the inequality from Theorem~\ref{th:mainLP} would be 
either $-\gamma \geq (1-\gamma)\sum_{i \in [k]}{\lambda_i}$ or $0 \geq (1-\gamma)\sum_{i \in [k]}{\lambda_i}$;
but those are not possible since the right-hand side is strictly positive by the previous item and the fact that $\gamma < 1$.
Thus, $X_0Y_0\subseteq \overline{X_1Y_1...X_kY_k}$.

{(3)} Because of the previous point, $(1-\gamma)\geq (1-\gamma) \cdot \sum_{i \in [k]}{\lambda_i}$, 
hence $1\geq \sum_{i \in \left[k\right]}{\lambda_i}$.

{(4)}, {(5)} and {(6)} Let $Z=\overline{X_0}$, by idempotency 
$Z = \overline{Z}$. Since $Y_0\nsubseteq \overline{X_0}$, we have $Y_0\nsubseteq X_0$ so $w_Z(X_0\rightarrow Y_0) = -\gamma$. From Theorem~\ref{th:mainLP} and 
Lemma~\ref{lm:mainLPBase}{(1)}, $-\gamma \geq -\gamma \cdot \sum_{i \in V_Z}{\lambda_i}$. Then $\sum_{i \in V_Z}{\lambda_i}\geq 1$ since $\gamma > 0$. 
By {(3)}, we have $\sum_{i \in \left[k\right]}{\lambda_i} \leq 1$,
and all are strictly positive; the only way to add up to 1 is $V_Z=\left[k\right]$.
Hence, $Z=\overline{X_0}$ violates every $X_i\rightarrow Y_i$, i.e., $X_i\subseteq \overline{X_0}$ and $X_iY_i \nsubseteq \overline{X_0}$; 
and, besides, $\sum_{i\in \left[k\right]}{\lambda_i}=1$.
 
{(7)} Let $Z = \overline{X_0Y_i}$ for any $i \in [k]$, again by idempotency 
$Z = \overline{Z}$. By {(4)} we get $X_i\subseteq \overline{X_0}$ and then 
$X_iY_i \subseteq \overline{X_0Y_i}$, that is, $i \in W_Z$. Let us assume $Y_0 \nsubseteq Z$ so $w_Z(X_0\rightarrow Y_0)=-\gamma$. 
We would have, from 
Theorem~\ref{th:mainLP} and
Lemma~\ref{lm:mainLPBase}{(3)}, 
$-\gamma \geq \lambda_i - \gamma \sum_{i \in U_Z}{\lambda_i} = \lambda_i - \gamma \sum_{i \in[k]}{\lambda_i}$ 
since by {(4)} $X_i \subseteq \overline{X_0} \subseteq Z$ for all $i \in [k]$. But this cannot be the case:
$\lambda_i - \gamma \sum_{i \in [k]}{\lambda_i}$ 
is strictly larger than $-\gamma$ because $\lambda_i > 0$ by {(1)} while $\sum_{i \in \left[k\right]}{\lambda_i} = 1$ by {(6)}. This contradiction proves that the assumption $Y_0 \nsubseteq Z$ was wrong. Thus $Y_0\subseteq Z = \overline{X_0Y_i}$
\qed

\section{Low thresholds: Cases of improper entailment}

As it turns out, there are some simple but interesting 
cases where our results allow us to prove that 
there cannot be any relative entailment 
as in Expression~\eqref{eqn:entailment} that does
not already hold from just one of its partial premises. 
The characterization, then, follows from known ones. 
This is what the two results of this section state. 
Both of them are very intuitive and might be known 
(although we have not been able to find any specific
reference).

In the first one, we discuss the case where the
antecedent of the conclusion is empty: this will
complement the picture when we discuss the
nonempty case below in Section~\ref{sec:middlethresholds}.
The other case is when
the confidence parameter $\g$ is too low.

We will need to apply the actual
variant of Theorem~\ref{th:case1} proved in \cite{Balcazar},
which differs from the one given above in that it allows for
the background set of classical implications~$\B$ and its
closure operator:

\begin{thmC}[\cite{Balcazar}] \label{th:case1withB}
Let $\gamma$ be a confidence parameter in $(0,1)$,
let $\B$ be a set of implications,
and let $X_0 \to Y_0$
and $X_1 \to Y_1$ be two partial implications. Then the following are
equivalent:
\begin{enumerate} 
\item $\B$, $X_1 \to Y_1 \models_\gamma X_0 \to Y_0$,
\item either $Y_0 \subseteq \overline{X_0}$, or $X_1 \subseteq \overline{X_0}$ and $X_0Y_0 \subseteq \overline{X_1Y_1}$
\end{enumerate}
\end{thmC}

\subsection{Empty antecedent in the conclusion}

For one of our results in Section~\ref{sec:middlethresholds},
it will be useful to have studied separately
the case where $X_0 \subseteq \cls{\emptyset}$
(that is, $X_0 = \emptyset$ or equivalent to it under $\B$,
since $X_0 \subseteq \cls{\emptyset}$ if and only if $\cls{X_0} = \cls{\emptyset}$).

\begin{prop}
\label{prop:emptyantec}
Let $\gamma$ be a confidence parameter in (0,1), let $X_0 \to Y_0, \ldots , X_k \to Y_k$ be a set of partial implications, 
and let $\B$ be a set of implications such that
the entailment $\B, X_1 \to Y_1, \ldots, X_k \to Y_k \models_{\g} X_0 \to Y_0$ holds, 
where $X_0 \subseteq \cls{\emptyset}$.
Then, there is $j\in[k]$ such that 
$\B, X_j \to Y_j \models_{\g} X_0 \to Y_0$.
\end{prop}

To prove it, we apply Theorem~\ref{th:case1withB}:
the intended conclusion of this proposition is equivalent to:
either $Y_0 \subseteq \overline{X_0}$, or 
$X_j \subseteq \overline{X_0}$ and $X_0Y_0 \subseteq \overline{X_jY_j}$
for some $j\in[k]$;
note that this alternative formulation is again independent of $\g$.

\proof
Assume that the indicated entailment holds.
We start by passing to a subset for which the entailment
is proper as per Proposition~\ref{prop:properent}:
fix $L$ such that the entailment 
$\B,\{ X_i \to Y_i : i \in L \} \models_\gamma X_0 \to Y_0$ 
holds properly. 

If $L = \emptyset$, then any $j\in[k]$ will do, as we will have the
entailment already just from~$\B$ ($Y_0 \subseteq \overline{X_0}$).
Thus, assume $L$ nonempty, and fix any $j\in L$. 
By Lemma~\ref{prop:chaos}{(4)},
$X_j \subseteq \overline{X_0}$. 
Also, using monotonicity and idempotency of closures,
$X_0 \subseteq \cls{\emptyset} \subseteq \cls{Y_j}$
so that $X_0Y_j \subseteq \cls{Y_j}$;
then, 
by Lemma~\ref{prop:chaos}{(7)},
$Y_0 \subseteq \cls{X_0Y_j}\subseteq \cls{Y_j}$,
that is, considered together, $X_0Y_0 \subseteq \cls{Y_j} \subseteq \cls{X_jY_j}$
as well.
Our claim follows from Theorem~\ref{th:case1withB}. 
Note that, necessarily, $|L| = 1$ in this case.
\qed

\subsection{Low thresholds}

The remainder of the paper will be driven by a case study
based on the value of $\gamma$. First, we see that when it
is below a certain value, every entailment trivializes in the
same sense as the one just described in Proposition~\ref{prop:emptyantec}.
In the next section, we will study the case where it is 
high enough that the solution vector $\lambda$ can be chosen
to have the same value at all nonzero components, and 
another section will explain what happens for intermediate
values of $\gamma$. 

Our main result for low $\gamma$ is:

\begin{thm} \label{th:lowgamma}
Let $\gamma$ be a confidence parameter in $(0,1)$, let $X_0 \to Y_0,\ldots,X_k \to Y_k$ be a set of partial implications
with $k \geq 1$ and let $\B$ be a set of implications. If $\gamma < 1/k$, then the following are equivalent:
\begin{enumerate} 
\item $\B, X_1 \to Y_1,\ldots, X_k \to Y_k \models_\g X_0 \to Y_0$,
\item $\B, X_i \to Y_i \models_\g X_0 \to Y_0$ for some $i \in [k]$,
\item either $Y_0 \subseteq \overline{X_0}$, or $X_i \subseteq \overline{X_0}$ and $X_0Y_0
  \subseteq \overline{X_iY_i}$ for some $i \in [k]$.
\end{enumerate}
\end{thm}

\proof
The equivalence of {(2)} and {(3)} 
is exactly Theorem~\ref{th:case1withB}.
Also, {(2)}${}\Rightarrow{}${(1)} is
immediate.

To complete the proof we argue that
{(1)}${}\Rightarrow{}${(2)}. 
Let $L \subseteq [k]$ be minimal under set inclusion with 
$\B,\{ X_i \to Y_i : i \in L \} \models_\g X_0 \to Y_0$
as in Proposition~\ref{prop:properent}.
If $|L| \leq 1$ we already have what we want, because 
either $L = \emptyset$ and $\B \models_\gamma X_0 \to Y_0$,
that is, $Y_0 \subseteq \overline{X_0}$, or $|L| = 1$ and then,
for $i\in L$, 
$\B, X_i \to Y_i \models_\gamma X_0 \to Y_0$. 
Now, assuming $|L|\geq 2$, we just have to prove $\gamma \geq 1/k$,
thus contradicting our hypothesis.

Let $\l = (\l_i : i \in L)$ be a solution to 
Expression~\eqref{eqn:inequalities} for $\B, \{ X_i \to Y_i : i \in L \}
\models_\gamma X_0 \to Y_0$ as per Theorem~\ref{th:mainLP}. By the
minimality of $L$, that entailment 
is proper. As $\gamma$ is in $(0,1)$ and $|L|\geq 1$, indeed $|L|\geq 2$, Lemma~\ref{prop:chaos} 
applies, so we have $X_i \subseteq \overline{X_0}$ for all $i \in L$.
By the fact that $|L|\geq 2$, and the characterization of entailment with at most one premise, 
Theorem~\ref{th:case1withB},
we have $X_0Y_0 \nsubseteq \overline{X_iY_i}$ for all $i \in L$:
otherwise, the entailment would not be proper.
Taking a fixed $i$ in $L$, and $Z=\overline{X_iY_i}$, where $\overline{Z} = \overline{\overline{X_iY_i}} = \overline{X_iY_i} = Z$, we have:
$w_Z(X_0\to Y_0) \leq 0$, $w_Z(X_i \to Y_i) = 1-\g$. Since $W_Z \neq \emptyset$
and the entailment is proper, by Theorem~\ref{th:mainLP} and Lemma~\ref{lm:mainLPBase}{(3)}
we get
$0 \geq  \l_i - \g \cdot \sum_{j \in L} \l_j$. By Lemma~\ref{prop:chaos}{(3)}, we have
$\sum_{j \in L} \l_j \leq 1$ so $0 \geq \l_i - \g$. We conclude that $\l_i \leq \g$, and
this holds for every $i \in L$. Adding over $i \in L$ we get $\sum_{i
  \in L} \l_i \leq \g \cdot |L|$, and the left-hand side is~$1$ by
Lemma~\ref{prop:chaos}{(6)}. Thus $\g \geq 1/|L| \geq 1/k$ and the
theorem is proved.
%
%
%
%
\qed

It has been suggested to us that this result can be probably
obtained without resorting to the linear programming 
framework. That observation may be correct. However, the 
argument, as given, is an excellent way to start the study and
get a first contact with the usage of our toolkit.

\section{High thresholds} \label{sec:high} 

The goal of this section is to characterize entailments from $k$
partial implications when the confidence parameter $\gamma$ is
large enough, and our proofs will show that $(k-1)/k$ is
enough. Ideally, the characterization should make it easy to decide 
whether
an entailment holds, or at least easier than solving the linear
program given by Theorem~\ref{th:mainLP}. We come quite close to
that. Before we get into the characterization, let us first discuss
the key new concept on which it rests.

\begin{defi}
We say that a set of partial implications $X_1 \to Y_1,\ldots,X_k \to Y_k$ 
\emph{enforces homogeneous implicational satisfaction}
for a given set of implications $\B$ and its closure operator, 
if, for every $Z = \overline{Z}$, the following holds:
\begin{tabbing}
\indent\indent\indent
  \= \underline{if} \= for all $i \in [k]$ either $X_i \not\subseteq Z$ \underline{or} $X_iY_i \subseteq Z$ holds, \\
  \> \underline{then} \= either $X_i \not\subseteq Z$ holds for all $i \in [k]$ \\
  \> \> \underline{or}  $X_iY_i \subseteq Z$ holds for all $i \in [k]$.
\end{tabbing}
For economy of words, when the condition holds we
most often say that the set of partial implications
``enforces homogeneity'' for $\B$.
\end{defi}

In words, enforcing homogeneous implicational satisfaction
means that every $Z$ that does not
violate any $X_i \to Y_i$, either witnesses them all, or does not
cover any of them. Seen as the classical implication counterparts,
if they are all simultaneously satisfied, then they are satisfied
in a homogeneous manner: all of them vacuously, or all of them
witnessed. Note that this definition does not depend on any
confidence parameter. 

If $\B$ is not mentioned, we refer to $\B = \emptyset$,
with the trivial closure operator associated, namely the
identity; then, the condition defining homogeneous 
implicational satisfaction applies to every $Z$.

Sets of less than two
partial
implications 
always enforce homogeneity;
in the case of the empty set, 
it does so vacuously.
Thus, every set of partial implications has some 
subset that enforces homogeneity.

\jlb{For economy of words,
sometimes we refer to a set of partial implications that enforces
homogeneity as being \emph{nice}.
--- Trying to remove "nice"!}

\aa{Estaba demasiado cableado para hacerlo desaparecer.}

\jlb{Lo intento de todos modos. Sin embargo, me queda una duda.
El termino original era ``enforces homogeneous 
implicational satisfaction'' que es mucho mas claro y preciso
pero demasiado largo. Puestos a hacerlo corto, ``nice''
tampoco es mucho peor que ``enforces homogeneity''. 
No acabo de tener claro que hacer.}

\subsection{Enforcing homogeneity: main property} 

Homogeneity sounds like a very strong requirement. However, as the
following lemma shows, it is at the heart of proper entailments.
\begin{lem}
\label{lem:nicetynew}
Let $X_1 \to Y_1,\ldots,X_k \to Y_k$ be a set of partial implications
with $k \geq 1$ and let $\B$ be a set of implications. 
If there exist a partial implication $X_0 \to Y_0$ and a confidence 
parameter $\gamma$ in the interval $(0,1)$, for which the entailment 
$\B, X_1 \to Y_1,\ldots,X_k \to Y_k \models_\gamma X_0 \to Y_0$ 
holds properly, then
$X_1 \to Y_1,\ldots,X_k \to Y_k$ enforces
homogeneity for $\B$.
\end{lem}

\proof
By definition, we have to prove that if $\forall i \in \left[k\right]\,w_Z(X_i\rightarrow Y_i)\neq -\gamma$ then either $\forall i \in [k]\,w_Z(X_i\rightarrow Y_i) = 1-\gamma$ 
or $\forall i \in [k]\,w_Z(X_i\rightarrow Y_i) = 0$;
and this for all $Z$ such that $Z=\overline{Z}$.

Thus, fix such a $Z=\overline{Z}$ and assume indeed that $\forall i \in \left[k\right]$, $w_Z(X_i\rightarrow Y_i)\neq -\gamma$, that is,
for all $i\in [k]$,
either $w_Z(X_i\rightarrow Y_i) = 1-\gamma$ or $w_Z(X_i\rightarrow Y_i) = 0$. Thus, $w_Z(X_i\rightarrow Y_i) \geq 0$.
If $\forall i\in \left[k\right]\,w_Z(X_i\rightarrow Y_i) = 0$, then we are done.
Then, let us assume that there exists $j$ such that $w_Z(X_j\rightarrow Y_j) \neq 0$, so that 
$w_Z(X_j\rightarrow Y_j) = 1 - \gamma$, $X_jY_j \subseteq Z$, and $j\in W_Z$.
We apply Theorem~\ref{th:mainLP} with all weights and coefficients
non-negative, hence:
$$
w_Z(X_0\rightarrow Y_0) \geq \sum_{i=1}^k \lambda_i \cdot w_Z(X_i \rightarrow Y_i) \geq 
\lambda_j\cdot (1-\gamma) + \sum_{i\neq j} \lambda_i \cdot w_Z(X_i \rightarrow Y_i) > 0
$$
because $\lambda_j\cdot (1-\gamma) > 0$ already.
This implies $w_Z(X_0\rightarrow Y_0) = 1-\g$ so that 
$X_0\subseteq X_0Y_0\subseteq Z$ and therefore $\overline{X_0} \subseteq \overline{Z}=Z$ by monotonicity of closures.

By Lemma~\ref{prop:chaos}{(4)} we know 
$\forall i \in [k]\,X_i \subseteq \overline{X_0}$, so $X_i \subseteq Z$ and, 
by hypothesis, $Z$ does not violate any $X_i\rightarrow Y_i$; 
then $\forall i \in [k]\,X_iY_i\subseteq Z$, just what we want to prove. 
\qed

\subsection{Main result for high threshold}

We are ready to state and prove the characterization theorem for
$\gamma \geq (k-1)/k$.

\begin{thm}
\label{th:mainHGnew}
Let $\gamma$ be a confidence parameter in $(0,1)$, let $X_0 \to
Y_0,\ldots,X_k \to Y_k$ be a set of partial implications with $k \geq
1$ and let $\B$ be a set of implications. If $\gamma \geq (k-1)/k$ then the following are equivalent:
\begin{enumerate} 
\item $\B, X_1 \to Y_1,\ldots,X_k \to Y_k \models_\gamma X_0 \to Y_0$,
\item there is a set $L \subseteq [k]$ such that $\B,\{ 
  X_i \to Y_i : i \in L \} \models_\gamma X_0 \to Y_0$ holds properly,
\item either $Y_0 \subseteq \overline{X_0}$, or there is a non-empty $L
\subseteq [k]$ such that the following conditions hold:
\begin{enumerate} 
\item 
$\{ X_i \to Y_i : i \in L \}$ enforces homogeneity for $\B$,
\item 
$\bigcup_{i \in L} X_i \subseteq \overline{X_0} \subseteq \overline{\bigcup_{i \in L} X_iY_i}$,
\item 
$Y_0 \subseteq \bigcap_{i \in L} \overline{X_0Y_i}$.
\end{enumerate}
\end{enumerate}
\end{thm}

\proof
The equivalence of {(1)} and {(2)} is Proposition~\ref{prop:properent}.

From {(2)}~to {(3)}, the index set $L$ will be the same in
  both statements, unless $L = \emptyset$, in which case $Y_0
  \subseteq \overline{X_0}$ must hold and we are done. Assume then that $L$ is
  not empty. Part {(a)} we get automatically from
  Lemma~\ref{lem:nicetynew} since $\B, \{X_i \to Y_i : i \in L \}$
  properly entails $X_0 \to Y_0$ at threshold $\gamma$. 
  Now we prove {(b)}. By Theorem~\ref{th:mainLP}, let
  $\l = (\l_i : i \in L)$ be a solution to the inequalities
  in Expression~\eqref{eqn:inequalities} for the entailment $\B,\{X_i \to Y_i : i
  \in L\} \models_\gamma X_0 \to Y_0$. From the fact that this
  entailment is proper and the assumptions that $|L| \geq 1$ and
  $\gamma \in (0,1)$, we are allowed to apply Lemma~\ref{prop:chaos}.

  The first inclusion in {(b)} follows from that lemma, {(4)}.
  The second inclusion in {(b)} also follows from that lemma,
  {(2)}. Finally, for~{(c)} we just refer to {(7)} of the
  same lemma, where 
  we get $Y_0 \subseteq \overline{X_0Y_1},\ldots ,Y_0 \subseteq \overline{X_0Y_k}$
	and then $Y_0 \subseteq \bigcap_{i \in k} \overline{X_0Y_i}$
	(which is itself a closed set).

  For the implication from~{(3)}~to~{(1)}~we proceed as
  follows. If $Y_0 \subseteq \overline{X_0}$ then $X_0 \to Y_0$
  is already entailed by $\B$ (even $X_0 \dto Y_0$ is).
  Assume then that $L$ is non-empty and
  satisfies {(a)}, {(b)}, and {(c)}. By
  Theorem~\ref{th:mainLP} it suffices to show that the inequalities
  in Expression~\eqref{eqn:inequalities} for the entailment $\B, \{ X_i \to Y_i : i
  \in L \} \models_\gamma X_0 \to Y_0$ have a solution $\l = (\l_i : i
  \in L)$ with non-negative components.

  Let $\ell = |L|$ and set $\l_i = 1/\ell$ for $i\in L$. Recall that
  $L$ is not empty so $\ell \geq 1$ and this is well-defined. For
  fixed $Z$, we prove that the inequality in Expression~\eqref{eqn:inequalities}
  for this $Z$ is satisfied by these $\l_i$. If $Z \neq \overline{Z}$, 
  then that inequality 
is satisfied trivially 
  since all weights at both sides of the inequality are zero. Thus, let
$Z = \overline{Z}$. In the following, 
let $X = \bigcup_{i\in L} X_i$ and $Y = \bigcap_{i\in L} \overline{Y_i}$.  We
  distinguish cases according to whether $X\subseteq Z$.

First assume that $X\subseteq Z$. In this case $Z$ covers $X_i \to Y_i$ 
for every $i \in L$: then $L = U_Z$. 
Thus, we split $L$ into two sets, $L = V_Z \cup W_Z$. 
We consider three subcases.

\noindent
{\it Subcase~1}. If $W_Z = \emptyset$, then by Lemma~\ref{lm:mainLPBase}{(2)}, 
$\sum_{i \in L} \l_i \cdot w_Z(X_i \to Y_i) = -\g
\cdot \sum_{i \in L} \l_i$ and, using that the $\lambda_i$'s add up to 1, $ -\g
\cdot \sum_{i \in L} \l_i = -\g \leq w_Z(X_0 \to Y_0)$; i.e. the
inequality holds.

\noindent
{\it Subcase~2}. If $W_Z= L$, then every $X_i \to Y_i$ with $i \in L$ is witnessed:
$X_iY_i \subseteq Z$ for all $i\in L$.
Using {(b)}, and monotonicity and idempotency of the closure operator,
we get $\overline{X_0} \subseteq \overline{\bigcup_{i\in L}X_iY_i}
\subseteq \overline{Z} = Z$, and the non-emptiness of $L$ applied to {(c)}
ensures the existence of some $i \in L$ for which 
$Y_0\subseteq \overline{X_0Y_i}\subseteq \overline{Z} = Z$.
Thus $X_0 \to Y_0$ is also witnessed,
all the weights in the inequality \eqref{eqn:inequalities} are $1 - \g$,
the coefficients add up to 1, and
the inequality holds.

\noindent
{\it Subcase~3}. We consider now the general case where $W_Z \neq\emptyset$ and $W_Z \neq L$.
The fact that $W_Z \neq\emptyset$ ensures
that there is some $i \in L$ such that 
$Y_i \subseteq Z$. By {(c)}, $Y_0 \subseteq X_0Y_i$;
then, $w_Z(X_0 \to Y_0) \geq 0$, specifically $1-\g$ or $0$ 
according to whether 
$X_0 \subseteq Z$. 
The fact that $W_Z \neq L$ implies that $V_Z \neq \emptyset$;
by Lemma~\ref{lm:mainLPBase}{(4)}, with $U_Z = L$,
$\sum_{i \in L}{\lambda_i w_Z(X_i \to Y_i)} \leq 
\sum_{i \in L - \left\{ j \right\}}{\lambda_i}-\gamma\sum_{i \in L}{\lambda_i}$
where $j\in V_Z$. As all the $\lambda_i$ are $1/\ell$:
\begin{eqnarray*}
\sum_{i \in L}{\lambda_i w_Z(X_i \to Y_i)} \leq \frac{\ell -1}{\ell}- \gamma \leq \frac{k -1}{k}- \gamma \leq 0 \leq w_Z(X_0 \to Y_0).
\end{eqnarray*}


Assume now instead $X \not\subseteq Z$. Then, by the first inclusion in
{(b)}, $\overline{X_0} \not\subseteq Z$, so $Z$ does not cover $X_0 \to Y_0$
and $w_Z(X_0 \to Y_0) = 0$. 
If $\overline{X_iY_i} \not\subseteq Z$ for every $i\in L$, then $Z$ does not
witness any $X_i \to Y_i$, so $w_Z(X_i \to Y_i) \leq 0$ for every $i
\in L$. Whence $\sum_{i\in L} \l_i \cdot w_Z(X_i \to Y_i)$ is
non-positive and then bounded by $w_Z(X_0 \to Y_0) = 0$ as required.
Hence, suppose now that there exists $q \in L$ such that $X_qY_q
\subseteq Z$. As $X \not\subseteq Z$, we also have a $j \in L$ such that $X_j \not\subseteq
Z$.  Thus $Z$ witnesses $X_q \to Y_q$ and fails to cover $X_j \to
Y_j$, and both $q$ and $j$ are in $L$. As $\{X_i \to Y_i : i \in L
\}$ enforces homogeneity, this means that $Z$ must violate $X_h \to Y_h$
for some $h \in L$: $h\in V_Z \neq \emptyset$; 
then, by Lemma~\ref{lm:mainLPBase}{(4)} we have:
$$
\sum_{i \in L} \l_i  \cdot w_Z(X_i \to Y_i)\leq
\sum_{i \in U_Z - \left\{ h \right\}}{\lambda_i}-\gamma\sum_{i \in U_Z}{\lambda_i}.
$$ 

Let $\left| U_Z \right| = u < \ell \leq k$. Replacing the values of $\lambda_i$,
%
%
$$
\sum_{i \in L} \l_i  \cdot w_Z(X_i \to Y_i)\leq
\frac{1}{\ell}(u-1) - \frac{1}{\ell}u\g = \frac{u(1-\g) - 1}{\ell} \leq{} 
$$

$$
{} \leq\frac{\ell(1-\g) - 1}{\ell}
= \frac{\ell - 1}{\ell} - \g \leq \frac{k - 1}{k} - \g.
$$ 
In turn, this last value is non-positive, 
and thus bounded by $w_Z(X_0 \to Y_0) = 0$, by the
assumption that $\g \geq (k-1)/k$. This proves that the inequalities
corresponding to these $Z$'s are satisfied.

This closes the cycle of implications and the theorem is proved.
\qed



\subsection{Enforcing homogeneity: further properties} \label{sec:morenice}

Enforcing homogeneity 
turned out to play a key role in
the main result about the case of high confidence threshold. 
In this section we collect a few
additional observations about it. 

We already mentioned the case of sets 
of less than two partial implications.
The case $k = 2$ is a bit more interesting. 
We find that this case of enforced homogeneity
corresponds exactly
to the conditions under label (a) in 
Theorem~\ref{th:case2}(2),
that we found mysterious for many years (cf.~the
discussion at the end of Section~\ref{sec:uptotwopremises}).

\begin{lem}
A set of two partial implications $X_1 \to Y_1, X_2 \to Y_2$ enforces
homogeneity for $\B$ if and only if both $X_1\subseteq \overline{X_2Y_2}$ and
$X_2\subseteq \overline{X_1Y_1}$ hold.
\end{lem}
\proof
($\Rightarrow$) 
Pick $Z = \overline{X_2Y_2}$; then, assuming $X_1 \nsubseteq Z$
directly violates the definition of enforcing homogeneity; then
argue symmetrically.

($\Leftarrow$) Assume $X_1 \subseteq \overline{X_2Y_2}$ and $X_2 \subseteq \overline{X_1Y_1}$, 
and let $Z = \overline{Z}$.
Suppose that either $X_1Y_1 \subseteq Z$ or $X_1 \nsubseteq Z$, and likewise
either $X_2Y_2 \subseteq Z$ or $X_2 \nsubseteq Z$, but not homogeneously.
Rename if necessary so that $X_1 \nsubseteq Z$ and $X_2Y_2 \subseteq Z$,
and apply monotonicity and idempotency: $\overline{X_2Y_2} \subseteq \overline{Z} = Z$;
then, $X_1 \subseteq \overline{X_2Y_2}$ is not possible.
Thus, homogeneity holds.
\qed

The next lemma characterizes sets of partial
implications that enforce homogeneity, and provides
us with a polynomial-time algorithm to test them. 

\begin{lem}
\label{lm:antecsnicenew}
Let $X_1 \to Y_1,\ldots,X_k \to Y_k$ be a set of partial implications
and let $U= X_1Y_1 \cdots X_kY_k$. Then, the following are
equivalent:
\begin{enumerate} 
\item $X_1 \to Y_1,\ldots,X_k \to Y_k$ enforces homogeneity for $\B$,
\item $\B, X_1 \Rightarrow Y_1,\ldots,X_k \Rightarrow Y_k \models
X_i \Rightarrow U$, all $i \in [k]$.
\end{enumerate}
\end{lem}

\proof
  Assume $X_1\to Y_1,\ldots,X_k \to Y_k$ enforces homogeneity for $\B$; 
  let $Z\models\B$, that is, $Z = \overline{Z}$, and assume that
  $Z\models X_i\Rightarrow Y_i$ for all $i\in [k]$. Then, by
  homogeneity, either $X_i\not\subseteq Z$ for all $i \in [k]$ or $X_iY_i\subseteq Z$ for all $i\in [k]$.
	If $X_i\not\subseteq Z$ for all $i\in [k]$, then it also holds $Z\models X_i\Rightarrow U$ for all $i\in [k]$.
  Else, if $X_iY_i\subseteq Z$ for all $i\in [k]$ then $U\subseteq Z$,
  and $Z\models X_i\Rightarrow U$ for all $i\in [k]$ as
  well. Therefore, $X_1\Rightarrow Y_1,\ldots,X_k\Rightarrow Y_k$
  entail every $X_i\Rightarrow U$.

  Conversely, assume that $\B, X_1 \Rightarrow Y_1,\ldots,X_k \Rightarrow Y_k$ 
  entail every $X_i\Rightarrow U$ and let $\overline{Z}=Z\models X_i\Rightarrow Y_i$ 
  for all $i\in [k]$, hence	$Z\models X_i\Rightarrow U$ for all $i\in [k]$. 
	Then either $U\subseteq Z$, in this case $X_iY_i \subseteq Z$ for all $i \in [k]$ and we are done;
	or $U\not\subseteq Z$ and, then, the only way to satisfy all these
  classical implications is by falsifying all the premises, so that
  $X_i\not\subseteq Z$ for all $i\in [k]$. 
	Therefore we have proved that $X_1 \to Y_1,\ldots,X_k \to Y_k$ enforces homogeneity.
\qed

Note that condition~{(2)} in the
lemma can be decided efficiently by testing the unsatisfiability of
all the propositional Horn formulas of the form $\bigwedge\B \wedge (X_1 \Rightarrow Y_1)
\wedge \cdots \wedge (X_k \Rightarrow Y_k) \wedge X_j \wedge \neg A$
as $j$ ranges over $[k]$ and $A$ ranges over the attributes in $U$.

This characterization is quite useful. Consider, for instance, the set of three
partial implications $B \to ACE, C \to AD, D \to AB$ on the attributes
$A,B,C,D,E$. By the lemma, this set enforces homogeneity, but any of
its two-element subsets fails to do so. 

Finally, a
recurrent situation concerns sets of partial implications
with a common left-hand side; more generally, when the closures
of the left hand sides coincide.






\begin{lem}
Every set of partial implications of the form $X_1 \to Y_1,\ldots,X_k \to Y_k$ 
such that, for some $X$ and all $i \in [k]$, $\overline{X_i} = X$, enforces homogeneity for $\B$.
\end{lem}

\proof
Consider any $Z = \overline{Z}$ such that 
for all $i \in [k]$ either $X_i \nsubseteq Z$ or $X_iY_i \subseteq Z$.
If $X_i \nsubseteq Z$ for all $i \in [k]$, then homogeneity
is enforced. Assume that, for some $j \in [k]$, $X_j \subseteq Z$;
then $X_i \subseteq Z$ for all $i \in [k]$, since
$X_i \subseteq \overline{X_i} = \overline{X_j} \subseteq \overline{Z} = Z$,
and the only remaining option is that $X_iY_i \subseteq Z$ for all $i \in [k]$,
again enforcing homogeneity.
\qed

%

\section{Intervening thresholds} \label{sec:middlethresholds}

The rest of the values of $\gamma$ require \emph{ad hoc} consideration in
terms of the actual partial implications involved. We start by
defining what will end up being the \emph{critical} confidence
threshold for a given entailment.

\aa{Chronologically, the material of this section came
  first. The point of this characterization is that the necessary
  conditions it states are very informative. Moreover, the special
  case $\gamma \geq (k-1)/k$ provides a purely set-theoretic
  characterization (because $\gamma^* \leq (|K|-1)/|K|$; but why was
  that?).} 

\jlb{No estaría mal repasar esa pregunta...}


\begin{defi}
Let $\Sigma = \{ X_1 \to Y_1,\ldots,X_k \to Y_k \}$ be a set of
partial implications with $k \geq 1$, let $\B$ be a set of implications and all their attributes in $[n]$,
and let $X \subseteq [n]$ with $X\neq\emptyset$. Define
the \emph{critical threshold} for $\Sigma$ and $X$ as follows:
\begin{equation}
\gamma^* = \gamma^*(\Sigma, X) := \inf_{\lambda} \max_Z
\frac{\sum_{i \in W_Z} \lambda_i}{\sum_{i \in U_Z} \lambda_i}
\label{eqn:gammastar}
\end{equation}
where
$Z$ ranges over all subsets of $[n]$ with $X \not\subseteq Z$ and $Z = \overline{Z}$
according to $\B$, and
$\lambda$ ranges over vectors $(\l_1,\ldots,\l_k)$ of
  non-negative reals such that $\sum_{i \in [k]} \lambda_i = 1$.
Moreover, by convention, in Expression~\eqref{eqn:gammastar},
any occurrence of $0/0$ 
is taken as $0$.
\end{defi}

We can also agree that a vacuous maximum is taken as $0$;
however, note that this last case occurs only if $X \subseteq\cls{\emptyset}$ since
otherwise there is always the possibility of taking $Z = \cls{\emptyset}$.
We will avoid usage of this definition for these $X$ as that
case was already covered in Propositon~\ref{prop:emptyantec}.
On the other hand, we required $k \geq 1$. This ensures
that the $\inf$ is not vacuous, which in turn implies $0 \leq \gamma^*
\leq 1$: the lower bound is obvious, and for the upper bound just take
$\l_i = 1/k$ for every $i\in[k]$, which is well-defined when $k \geq 1$.

Observe that $\gamma^*$ is defined for a set of partial implications
and a single set $X$ of attributes. Typically $X$ will be the
left-hand side of another partial implication $X_0 \to Y_0$, but
$\gamma^*(\Sigma,X_0)$ is explicitly defined not to depend on
$Y_0$. 

It should be pointed out that the convention about $0/0$ is \emph{not} an
attempt to repair a discontinuity; in general, the discontinuities of
the rational functions inside the max are not repairable. 
\jlb{en algun sitio teníamos algo mas dicho sobre eso...}
However, since all $\l_i$ are non-negative, the only way the
denominator can be zero is by making the numerator also zero;
jointly with our convention about 0/0, we will be able to avoid 
the fraction in the next proposition.

The bounds on $\lambda$ define a closed and bounded polytope;
thus, it is a compact set. It follows that the limit $\g^*$
is actually reached:

\aa{Should this be defined as $\inf$? Does it matter? Probably not.}
\aa{De hecho, ya no se porque con tantas convenciones ya no se si las
  cosas se extienden continuamente o que. Lo defino $\inf$ por si
  acaso. Esto obliga a $\epsilon$$\delta$'s en la demostracion.}
\aa{Por cierto, antes pedia $\l_i > 0$ y no lo usaba. Ahora solo pido
  $\l_i \geq 0$, lo cual hace que la minimizacion sea sobre un cerrado
  y tenga mejor pinta. Aun y asi... $\inf$ por si acaso.}

\jlb{A ver... creo que ya entiendo lo que ocurre, aunque ha pasado
un tiempo. En algun momento hubo una version con lambdas 
estrictamente positivos. Entonces el politopo es abierto, y
fuerza al juego de epsilones y deltas. Pero si no se usa que
los lambdas no sean nulos, y podemos admitir desigualdad no
estricta, entonces el politopo es un cerrado y acotado de $R^n$
lo cual implica compacto, lo cual implica que los extremos
se alcanzan... para funciones continuas. En nuestro caso
seguramente lo son pero, como los $Z$ van dando saltos 
discretos, es precisa una demo adicional de continuidad. Por otra parte,
la demostracion que tenemos en LICS no me vale, precisamente
por ese mismo motivo de que $Z$ no es estable. Se puede reparar
de una manera que me satisface, pero por casi el mismo precio
se puede demostrar que existe el $\lambda^*$ que alcanza 
el valor de $\g^*$ y prefiero hacerlo asi.}

\begin{prop} 
\label{prop:lambdastar}
Fix $\Sigma$ and $X$ as in the definition of 
$\gamma^* = \gamma^*(\Sigma, X)$
(with the same conventions).
Then there is a vector $\lambda^*$ such that, for every $Z$
such that $X \not\subseteq Z$ and $Z = \overline{Z}$,
$\frac{\sum_{i \in W_Z} \lambda_i}{\sum_{i \in U_Z} \lambda_i}\leq\g^*$;
equivalently,
$\sum_{i \in W_Z} \lambda^*_i \leq \g^*\sum_{i \in U_Z} \lambda^*_i$.
\end{prop} 

\jlb{This proof is quite standard basic Calculus, but I still
think it is necessary as the issue is delicate. If it breaks too much
the flow, it can be postponed to an appendix.}

\proof
For every non-negative integer $n$, we know that there is a vector
$\lambda^{(n)} = (\lambda^{(n)}_1,\ldots,\lambda^{(n)}_k)$ 
such that, for every $Z$
such that $X \not\subseteq Z$ and $Z = \overline{Z}$,
$\frac{\sum_{i \in W_Z} \lambda^{(n)}_i}{\sum_{i \in U_Z} \lambda^{(n)}_i} \leq\g^* + \frac{1}{n}$.
Taking into account our convention about 0/0, this is the same as
$\sum_{i \in W_Z} \lambda^{(n)}_i \leq (\g^* + \frac{1}{n}) \sum_{i \in U_Z} \lambda^{(n)}_i$
(recall that if the sum on $U_Z$ is zero, so is the sum on $W_Z \subseteq U_Z$).

We can rewrite that bound as follows:
$\sum_{i \in W_Z} \lambda^{(n)}_i \leq \g^* \sum_{i \in U_Z} \lambda^{(n)}_i + \frac{1}{n} \sum_{i \in U_Z} \lambda^{(n)}_i
\leq \g^* \sum_{i \in U_Z} \lambda^{(n)}_i + \frac{1}{n}$
given that a sum of $\lambda^{(n)}_i$'s is always bounded above by 1.
Thus, 
$\sum_{i \in W_Z} \lambda^{(n)}_i - \g^* \sum_{i \in U_Z} \lambda^{(n)}_i \leq \frac{1}{n}$.

The sequence $\{ \lambda^{(n)} \}$ must have at least one 
accumulation point 
$\lambda^* = (\lambda^*_1,\ldots,\lambda^*_k)$ in the polytope,
due to compactness.
We prove that
it enjoys the property as claimed. We argue the contrapositive,
by assuming that, for some $Z$,
$\sum_{i \in W_Z} \lambda^*_i > \g^*\sum_{i \in U_Z} \lambda^*_i$
or, equivalently, that $\eta > 0$ where
$\eta = \sum_{i \in W_Z} \lambda^*_i - \g^*\sum_{i \in U_Z} \lambda^*_i$
where, of course, $X \not\subseteq Z$ and $Z = \overline{Z}$.
Fix that $Z$ for the rest of the argument.

Let $n_0 > 2/\eta$, so that $1/n < \eta/2$ for every $n \geq n_0$
thanks to the assumption that $\eta > 0$.
Let $\delta = \frac{\eta}{2k(1+\g^*)}$ so that $\delta k(1+\g^*) = \eta/2$,
and let $n > n_0$ be large enough so that, for every $i\in [k]$,
$|\lambda^{(n)}_i - \lambda^*_i| \leq \delta$.

Then:
\begin{eqnarray*}
\eta &   =  & \sum_{i \in W_Z} \lambda^*_i - \g^*\sum_{i \in U_Z} \lambda^*_i \\
     & \leq & \sum_{i \in W_Z} (\lambda^{(n)}_i + \delta) - \g^*\sum_{i \in U_Z} (\lambda^{(n)}_i - \delta) \\
     &   =  & \sum_{i \in W_Z} \lambda^{(n)}_i - \g^*\sum_{i \in U_Z} \lambda^{(n)}_i + \delta(|W_Z| + \g^*|U_Z|) \\
     & \leq & \frac{1}{n} + \delta k(1 + \g^*) < \eta
\end{eqnarray*}
where the last two inequalities come directly from the properties
of $\lambda^{(n)}$, $n_0$, and $\delta$, and lead to a clearly
contradictory outcome. Thus, $\eta\leq 0$, and the claimed
property follows.
\qed

\jlb{Quito esto, es el lema nuevo.
For later reference let us also point out that, with the
notation $V_Z$ and $W_Z$ from above, the inequalities
in~\eqref{eqn:inequalities} for an entailment $X_1 \to Y_1,\ldots,X_k
\to Y_k \models_\g X_0 \to Y_0$ can be written as $w_Z(X_0 \to Y_0)
\geq (1-\g) \cdot \sum_{i \in W_Z} \l_i - \g \cdot \sum_{i \in V_Z}
\l_i$. It is not the first time we use this sort of notation.
}

\jlb{REVISAR LOS COMENTARIOS SIGUIENTES}

\aa{In the final version of the characterization theorem we should
  remove the hypothesis $Y' \not\subseteq X'$ and handle its failure
  by allowing the possibility that $|K|=0$.}  \aa{This is taken care
  by an escape clause ``Either $Y_0 \subseteq X_0$ or ...''.}
\aa{Initially I had allowed the case $k = 0$ but then I realized that
  the definition of $\gamma^*$ requires $k \geq 1$: otherwise the set
  of vectors $\l$ that add up to one is empty.}

\jlb{The proof also needs $X'\neq\emptyset$ because at some point we
  pick any $Z$ such that $X'\not\subseteq Z$. The particular case
  $X'=\emptyset$ probably works as well but, as of today, seems to
  require its own ad hoc proof.}  \aa{This has been taken care of by
  conveying that $\max \emptyset = 0$ in the definition of $\gamma^*$;
  is this good? I added a comment after the definition and in the
  proof.}
\jlb{Yes, but I think it would be helpful and instructive
to spell out what happens exactly for $X = \emptyset$.}

\subsection{Characterization for all thresholds}

The main result of this section is a characterization theorem in the
style of Theorem~\ref{th:mainHGnew} that captures all possible
confidence parameters.
  
\begin{thm}
\label{th:everygamma}
Let $\gamma$ be a confidence parameter in $(0,1)$, let $X_0 \to
Y_0, \ldots,X_k \to Y_k$ be a set of partial implications with $k \geq
1$ and let $\B$ be a set of implications. The following are equivalent:
\begin{enumerate} 
\item $\B, X_1 \to Y_1,\ldots,X_k \to Y_k \models_\gamma X_0 \to Y_0$,
\item there is a set $L \subseteq [k]$ such that $\B, \{X_i \to Y_i : i \in L \} \models_\g X_0 \to Y_0$ holds properly,
\item either $Y_0 \subseteq \overline{X_0}$, or there is a non-empty $L \subseteq
  [k]$ such that the following conditions hold:
\begin{enumerate} 
\item
$\{ X_i \rightarrow Y_i : i \in L \}$ enforces homogeneity for $\B$,
\item
$\bigcup_{i \in L} X_i \subseteq \overline{X_0} \subseteq \overline{\bigcup_{i \in L} X_iY_i}$,
\item
$Y_0 \subseteq \bigcap_{i \in L} \overline{X_0Y_i}$,
\item
either $\overline{X_0} = \cls{\emptyset}$ or
$\gamma \geq \gamma^*(\{ X_i \rightarrow Y_i : i \in L \},\overline{X_0})$.
\end{enumerate}
\end{enumerate}
\end{thm}

\noindent
Note that the case $\overline{X_0} = \cls{\emptyset}$, 
mentioned in {(d)}, trivializes
to $|L|\leq 1$, as proved in Proposition~\ref{prop:emptyantec}.

\proof
As before,
the equivalence of {(1)} and {(2)} is Proposition~\ref{prop:properent}.

\noindent
{(2)}${}\Rightarrow{}${(3)}
If $L = \emptyset$ then the entailment follows from $\B$ and
$Y_0 \subseteq \overline{X_0}$. 
Assume then that $L$ is not empty: part {(a)} follows from Lemma~\ref{lem:nicetynew}. 
By Theorem~\ref{th:mainLP}, we know that there exist $\lambda = (\lambda_i), i \in L$ solution to the inequalities in 
Expression~\ref{eqn:inequalities}. 
We apply Lemma~\ref{prop:chaos}.

By Lemma~\ref{prop:chaos}{(2)}, we have that $X_0\subseteq X_0Y_0 \subseteq  \overline{\bigcup X_iY_i}$. 
Thus, $\overline{X_0} \subseteq \overline{\bigcup X_iY_i}$ by monotonicity
and idempotency of the closure operator. The other inclusion in {(b)} follows from
Lemma~\ref{prop:chaos}{(4)}.

From Lemma~\ref{prop:chaos}{(7)}, we have $Y_0 \subseteq \overline{X_0Y_1},Y_0 \subseteq \overline{X_0Y_2},...$ 
so on for every $i \in L$ thus implies $Y_0 \subseteq \bigcap_{i \in L} \overline{X_0Y_i}$. 
This gives us {(c)}.

\jlb{up to here, this part is also exactly as before}

Let us prove {(d)}. First, note that for every $Z$ such that $X_0 \nsubseteq Z$ 
we have $w_Z(X_0\rightarrow Y_0)=0$
(although only those where $Z = \cls{Z}$ are relevant in the maximization for $\g^*$).
By Lemma~\ref{lm:mainLPBase}{(2)}, the corresponding inequality reads
$0 \geq \sum_{i \in W_Z}{\lambda_i} - \gamma \cdot\sum_{i \in U_Z}{\lambda_i}$.
Rearranging, we get $\gamma \geq (\sum_{i \in W_Z}{\lambda_i})/(\sum_{i \in U_Z}{\lambda_i})$
(in the general case) and the maximum of the right-hand side is $\gamma^*$, and thus $\gamma^*$ 
is also bounded by $\gamma$. 
Note that we get the same result for the particular case
of a null denominator because of how it is handled in
the definition of $\gamma^*$.

\noindent
{(3)}${}\Rightarrow{}${(1)}
If $Y_0 \subseteq \overline{X_0}$, 
then $\B, X_1 \to Y_1, \ldots, X_k \to Y_k \models_\gamma X_0 \to Y_0$ holds trivially. 
Assume $L$ non-empty; to prove $\B, X_1 \to Y_1, \ldots X_k \to Y_k \models_\gamma X_0 \to Y_0$ 
it is enough to find a solution to the inequality 
$\sum_{i \in L}{\lambda_i}\cdot w_Z(X_i \rightarrow Y_i)\leq w_Z(X_0 \rightarrow Y_0)$ for every $Z \in [n]$. 
If $Z \neq \overline{Z}$ the inequality is satisfied since all the weights are zero. 
Thus, fix $Z = \overline{Z}$; we prove $\sum_{i \in L}{\lambda_i}\cdot w_Z(X_i \rightarrow Y_i)\leq w_Z(X_0 \rightarrow Y_0)$ by cases.
\begin{enumerate}
\item First assume that $X_0Y_0 \subseteq Z$. Then, $Z$ witnesses $X_0\rightarrow Y_0$ and $w_Z(X_0\rightarrow Y_0) = 1 - \gamma$. 
By Lemma~\ref{lm:mainLPBase}{(1)}, the left-hand side can be written as 
$(1-\gamma)\cdot \sum_{i \in W_Z}{\lambda_i}-\gamma \cdot \sum_{i \in V_Z} \lambda_i$. 
Any solution with $\lambda_i \geq 0$ and $\sum_{i \in L}\lambda_i=1$
satisfies the inequality.

\item Now, we assume that $\overline{X_0} \subseteq Z$ but $Y_0 \nsubseteq Z$.
(As $Z$ is closed, this includes the case where $X_0 = \emptyset$.)
Since $X_0 \subseteq \overline{X_0}$, we have $w_Z(X_0\rightarrow Y_0)= -\gamma$. 
By {(b)} we have $X_i \subseteq \overline{X_0}$, whereas, by {(c)}, we know that 
$Y_0 \subseteq \overline{X_0Y_i}$ for every $i \in L$. 
Since $\overline{X_0} \subseteq Z$ and $Y_0 \nsubseteq Z$, this means that 
$X_i \subseteq Z$ but $Y_i \nsubseteq Z$ for every $i \in L$. It follows 
that $Z$ violates all the $X_i\rightarrow Y_i$ so that $w_Z(X_i \rightarrow Y_i) = - \gamma$ 
for every $i \in L$. 
Pick again any solution with $\lambda_i \geq 0$ and $\sum_{i \in L}\lambda_i=1$:
then the left-hand side of the inequality is $-\gamma \cdot \sum_{i \in L}{\lambda_i}=-\gamma=w_Z(X_0\rightarrow Y_0)$
so that
the inequality holds (with equality in this case). 

\item Assume $Z$ does not cover $X_0\rightarrow Y_0$, the
only remaining case: then $w_Z(X_0\rightarrow Y_0) = 0$.
Let $\lambda^* = (\lambda^*_i: i \in L)$ be
a vector attaining $\g^*$ as in Proposition~\ref{prop:lambdastar}:
for every closed $Z$ not including $X_0$,
$\sum_{i \in W_Z} \lambda^*_i \leq \g^*\sum_{i \in U_Z} \lambda^*_i$.
This last inequality can be rewritten as
$\sum_{i \in W_Z} \lambda^*_i - \g^*\sum_{i \in U_Z} \lambda^*_i \leq 0
= w_Z(X_0\rightarrow Y_0)$. The desired inequality follows
once more from Lemma~\ref{lm:mainLPBase}{(2)}.

\jlb{(The old argumentation with the epsilon here --- careful,
in case it is fished out back in, it needs to be reviewed with
utmost care and probably rewritten; but I hope to erase it
instead at some point in the coming weeks.)
Fix then a positive real $\epsilon > 0$ 
and let $\lambda = (\lambda_i: i \in L)$ 
be such that the max in the definition of 
$\gamma*$ is at most $\gamma* + \epsilon$. 
By (d) it is at most $\gamma + \epsilon$. 
So, we get 
$\gamma + \epsilon \cdot \sum_{i \in V_Z \cup W_Z}{\lambda_i} \geq \sum_{i \in W_Z}{\lambda_i}$ 
by non-negativity of $\lambda_i$. Rearranging, 
$(1-\gamma ) \cdot \sum_{i \in W_Z}{\lambda_i} -\gamma \cdot \sum_{i \in V_Z}{\lambda_i} 
\leq \epsilon \cdot \sum_{i \in V_Z \cup W_Z}{\lambda_i}$. Since $\lambda_i\geq 0$ 
and $\sum_{i \in L}{\lambda_i}\leq 1$, 
the right-hand side is at most $\epsilon$, which is 
precisely $w_Z(X_0 \rightarrow Y_0) + \epsilon$ 
since $Z$ does not cover $X_0 \rightarrow Y_0$ and 
$w_Z(X_0\rightarrow Y_0)=0$. 
This is the right-hand side in 
$\sum_{i \in L}{\lambda_i}w_Z(X_i\rightarrow Y_i) \leq w_Z(X_0\rightarrow Y_0) + \epsilon$; the inequality holds.}
\end{enumerate}

This closes the cycle of implications and the proof.
\qed

\aa{Esteticamente, esos ``enumerates'' no me gustan mucho. Acabare por
  quitarlos.}
\aa{Los he quitado porque ademas en el ieee-format no van ni pa tras.}

\jlb{como antes, veamos cómo los pone el formato LMCS}

\aa{I can't recall the proof of the thing below [that always $\gamma^*
  \leq (k-1)/k$, now gone].}

\jlb{Pero si logramos recuperar esta proposicion, yo la veo interesante:
  Always $\gamma^* \leq (k-1)/k$; hence, Theorem~\ref{th:mainHGnew}
  follows from Theorem~\ref{th:everygamma}.
}

\aa{Now that I think if it, maybe I just knew that $\gamma^* \leq
  (k-1)/k$ \underline{BY} Theorem~\ref{th:mainHGnew}. So the corollary
  goes backwards :-). But I am still confused: what if $\Sigma$ does
  not entail anything of the form $X \to Y$ with $Y \not\subseteq X$
  at any $\gamma$ at all?.}

\jlb{Connection to the case of low gamma? e.g. maybe via an upper
bound on gamma-star?}

\subsection{An interesting example}

\aa{He adaptado lo que habia a la definicion de $\gamma^*$ sin los
  problemas de indeterminaciones. La original estaba escrita con la
  otra. Lo he repasado (habia un typo en la transcripcion de la
  solucion numerica), y he anadido los comentarios del final).}

In view of the characterization theorems obtained so far, one may
wonder if the critical $\gamma$ of any entailment among partial
implications is of the form $(k-1)/k$. This was certainly the case for
$k = 1$ and $k = 2$, and Theorems~\ref{th:mainHGnew}
and~\ref{th:everygamma} may sound as hints that this could be the
case. In this section we refute this for $k = 3$ in a strong way: we
compute $\gamma^*$ for a specific entailment for $k = 3$ to find out
that it is the unique real solution of the equation
\begin{equation}
1-\gamma + (1-\gamma)^2/\gamma + (1-\gamma)^3/\gamma^2 = 1.
\label{eqn:equation}
\end{equation}
Numerically \cite{WolframAlpha}, the unique real solution is
$$
\gamma_{c} \approx 0.56984\ldots.
$$ 

\begin{exa}
Consider the following 5-attribute entailment for a generic confidence
parameter $\gamma$:
$$ 
B \rightarrow ACH,\; C \rightarrow AD,\; D \rightarrow AB
\;\models_\gamma\; BCDH \rightarrow A.
$$ 
Let us compute its $\gamma^*(\Sigma,X)$ where $\Sigma$ is the
left-hand side, and $X = BCDH$. In other words, we want to determine a
triple $\lambda = (\lambda_1,\lambda_2,\lambda_3)$ that minimizes
$$ 
\max_Z \frac{\sum_{i \in W_Z} \lambda_i}{\sum_{i \in V_Z \cup W_Z}
  \lambda_i}
$$ 
as $Z$ ranges over the sets that do not include $X = BCDH$, and
subject to the constraints that $\lambda_1,\lambda_2,\lambda_3 \geq 0$
and $\lambda_1 + \lambda_2 + \lambda_3 = 1$. There are $2^5 = 32$
possible $Z$'s out of which two ($ABCDH$ and $BCDH$) contain $X$ and
therefore do not contribute to the maximum. Some others give value $0$
to the ratio and therefore do not contribute to the maximum
either. Note that if either $|Z|\leq 2$, or $|Z|=3$ and $A \not\in Z$,
then $W_Z = \emptyset$, so the numerator is $0$ and hence the ratio is
also $0$ (recall the convention that $0/0$ is $0$). Thus, the only
sets $Z$ that can contribute non-trivially to the maximum are those of
cardinality $4$ or $3$ that contain the attribute $A$. There are four
$Z$ of the first type ($ABCD$, $ABCH$, $ABDH$ and $ACDH$) and six $Z$
of the second type ($ABC$, $ABD$, $ABH$, $ACD$, $ACH$ and $ADH$). The
corresponding ratios are
\begin{align*}
\frac{\lambda_2 + \lambda_3}{\lambda_1 + \lambda_2 + \lambda_3},
\frac{\lambda_1}{\lambda_1 + \lambda_2},
\frac{\lambda_3}{\lambda_1 + \lambda_3},
\frac{\lambda_2}{\lambda_2 + \lambda_3}, 
\frac{0}{\lambda_1 + \lambda_2},
\frac{\lambda_3}{\lambda_1 + \lambda_3},
\frac{0}{\lambda_1},
\frac{\lambda_2}{\lambda_2 + \lambda_3},
\frac{0}{\lambda_2},
\frac{0}{\lambda_3}.
\end{align*}
Those with $0$ numerator cannot contribute to the maximum so, removing
those as well as duplicates, we are left with
$$
\frac{\lambda_2 + \lambda_3}{\lambda_1 + \lambda_2 + \lambda_3},
\frac{\lambda_1}{\lambda_1 + \lambda_2},
\frac{\lambda_3}{\lambda_1 + \lambda_3},
\frac{\lambda_2}{\lambda_2 + \lambda_3}.
$$
Since all $\l_i$ are non-negative, the first dominates the third and
we are left with three ratios:
\begin{equation}
\frac{\lambda_2 + \lambda_3}{\lambda_1 + \lambda_2 + \lambda_3},
\frac{\lambda_1}{\lambda_1 + \lambda_2},
\frac{\lambda_2}{\lambda_2 + \lambda_3}. \label{eqn:terms}
\end{equation}
We claim\aa{What comes next is an analytical proof of what we once did
  through 3D plots on my desktop. We could consider including the
  plots as well.} that a $\lambda_{c}$ that satisfies the
constraints and minimizes the maximum of the three terms in
Expression~\eqref{eqn:terms} is
$$
\begin{array}{rcl}
\lambda_{c,1} & = &  1-\gamma_c \\
\lambda_{c,2} & = &  (1-\gamma_c)^2 / \gamma_c \\
\lambda_{c,3} & = &  (1-\gamma_c)^3 / \gamma_c^2 
\end{array}
$$
where $\gamma_c$ is the unique real solution of the equation
in Expression~\eqref{eqn:equation}.
Clearly this choice of $\lambda_c$ satisfies the
constraints of non-negativity, and they add up to one precisely
because their sum is the left-hand side in Expression~\eqref{eqn:equation}.  By
plugging in, note also that this $\lambda_c$ makes all three terms in
Expression \eqref{eqn:terms} equal to $\gamma_c$; that is,
\begin{equation}
\frac{\lambda_{c,2} + \lambda_{c,3}}{\lambda_{c,1} + \lambda_{c,2} + 
\lambda_{c,3}} = 
\frac{\lambda_{c,1}}{\lambda_{c,1} + \lambda_{c,2}} =
\frac{\lambda_{c,2}}{\lambda_{c,2} + \lambda_{c,3}} = 
\gamma_c. \label{eqn:maximum}
\end{equation}
 
\jlb{In the next paragraph we had the interval $(0,1)$. I don't see
how that can be possible since it would imply the extremes at
0 and 1 respectively. Luckily, upon using this inequality below,
we only need a smaller interval.}

For later reference, let us note that the left-hand side of
Expression~\eqref{eqn:equation} is a strictly decreasing function of $\gamma$ in
the interval $(0,\gamma_c]$ (which can be seen by differentiating it, 
or simply by plotting it) and
therefore
\begin{equation}
1-\gamma_0 + (1-\gamma_0)^2/\gamma_0 + (1-\gamma_0)^3/\gamma_0^2 > 1
\label{eqn:inequation}
\end{equation}
whenever $0 < \gamma_0 < \gamma_c$.

In order to see that $\lambda_c$ minimizes the maximum of the three
terms in Expression~\eqref{eqn:terms} suppose for contradiction that $\lambda$
satisfies the constraints and achieves a smaller maximum, say $0 <
\gamma_0 < \gamma_c$. Since $\gamma_0$ is the maximum of the three
terms in Expression~\eqref{eqn:terms} we have
$$
\begin{array}{rcl}
\gamma_0 & \geq & (\lambda_2 + \lambda_3)/(\lambda_1 + \lambda_2 + \lambda_3) \\
\gamma_0 & \geq & \lambda_1 / (\lambda_1 + \lambda_2) \\
\gamma_0 & \geq & \lambda_2 / (\lambda_2 + \lambda_3).
\end{array}
$$
Using $\l_1,\l_2,\l_3 \geq 0$ and $\lambda_1 + \lambda_2 + \lambda_3 =
1$, and rearranging, we get
$$
\begin{array}{rcl}
  \lambda_1 & \geq & 1 - \gamma_0 \\
  \lambda_2 & \geq & \lambda_1 \cdot (1-\gamma_0)/\gamma_0 \geq (1-\gamma_0)^2/\gamma_0 \\
  \lambda_3 & \geq & \lambda_2 \cdot (1-\gamma_0)/\gamma_0 \geq (1-\gamma_0)^3/\gamma_0^2.
\end{array}
$$
Adding all three inequalities we get
$$
\lambda_1 + \lambda_2 + \lambda_3 \geq 1-\gamma_0 +
(1-\gamma_0)^2/\gamma_0 + (1-\gamma_0)^3/\gamma_0^2.
$$
But this is a contradiction: the left-hand side is $1$ since
$\lambda$ satisfies the constraints, and the right-hand side is
strictly bigger than $1$ by Expression~\eqref{eqn:inequation}. This proves the
claim.

Finally, this example also shows that for $\gamma$ midway through
$1/k$ and $(k-1)/k$, the vector solution to the inequalities
in~\eqref{eqn:inequalities} could be very non-uniform. In this example
with $\gamma = \gamma_c$, the solution is
$\lambda_c \approx (0.43016, 0.32472,
0.24512)$.  In contrast, for $\gamma \geq (k-1)/k$, the proof of
Theorem~\ref{th:mainHGnew} shows that it is always possible to take
$\l_i = 1/|L|$ for $i \in L$ and $\l_i = 0$ for $i\in[k]\setminus L$.
In this case, the vector $(\lambda_1,\lambda_2,\lambda_3) =
(1/3,1/3,1/3)$ works for $\gamma \geq 2/3$, but fails otherwise.  To
see that it fails when $\gamma < 2/3$, take the inequality for $Z =
ABCD$ in Expression~\eqref{eqn:inequalities}.

By the way, 
Theorem~\ref{th:everygamma}
tells us that $\gamma_c \approx 0.56984$ is the smallest confidence at
which this entailment holds: indeed,
it is easy to check that conditions {(a)}, {(b)}
and {(c)} hold for this example, and thus (d) characterizes
entailment.
\end{exa}

\section{Closing remarks} \label{sec:closing}

Our study gives a useful handle on entailments among partial or
probabilistic implications. The very last comment of the previous
section is a good illustration of its power. However, there are a few
questions that arose and were not fully answered by our work.

For the forthcoming discussion, let us take $\gamma = (k-1)/k$ for
concreteness.  The linear programming characterization in
Theorem~\ref{th:mainLP} gives an algorithm to decide if entailment
holds that is polynomial in $k$, the number of premises, but
exponential in $n$, the number of attributes. This is due to the
dimensions of the matrix that defines the dual LP: this is a $2^n
\times k$ matrix of rational numbers in the order of $1/k$ (for our
fixed $\gamma = (k-1)/k$).  On the other hand, the characterization
theorem in Theorem~\ref{th:mainHGnew} reverses the situation: there
the algorithm is polynomial in $n$ but exponential in $k$. In order to
see this, first note that condition {(a)} can be solved by
running $O(nk)$ Horn satisfiability tests of size $O(nk)$ each, as
discussed at the end of Section~\ref{sec:morenice}. Second, conditions
{(b)} and {(c)} are really straightforward to check if the
sets are given as bit-vectors, say. So far we spent time polynomial in
both $n$ and $k$ in checking the conditions of the
characterization. The exponential in $k$ blow-up comes, however, from
the need to \emph{pass} to a subset $L \subseteq [k]$, as potentially
there are $2^k$ many of those sets to check. It does show, however,
that the general problem in the case of $\gamma \geq (k-1)/k$ is in
NP. This does not seem to follow from the linear programming
characterization by itself, let alone the definition of
entailment. But is it NP-hard?  Or is there an algorithm that is
polynomial in both $k$ and $n$? One comment worth making is that an
efficient \emph{separation oracle} for the exponentially many
constraints in the LP of Theorem~\ref{th:mainLP} might well exist,
from which a polynomial-time algorithm would follow from the ellipsoid
method.

It is tempting to think that the search over subsets of $[k]$ can be
avoided when we start with a proper entailment. And indeed, this is
correct. However, we do not know if this gives a characterization of
proper entailment. In other words, we do not know if conditions
{(a)}, {(b)} and {(c)}, by themselves, guarantee proper
entailment. The proof of the direction~{(3)} to~{(1)} in
Theorem~\ref{th:mainHGnew} does not seem to give this, and we suspect
that they do not. If they did, we would get an algorithm to check for
proper entailment that is polynomial in both $n$ and $k$.

From a wider and less theoretical prespective, it would be very
interesting to find real-life situations in problems of data analysis,
say, in which partial implications abound, but many are redundant. In
such situations, our characterization and algorithmic results could
perhaps be useful for detecting and removing such redundancies, thus
producing outputs of better quality for the final user. This was one
of the original motivations for the work in \cite{Balcazar}, and our
continuation here. Also, again along the same lines, it has been 
argued that confidence, while the most natural measure for the
strength of a partial implication, may not be the most useful one
in practice, and a number of alternatives have been put forward
(see~\cite{GH}). Redundancy studies, like the one developed here
for confidence, are definitely worthwhile for the most commonly 
employed among these.

\bibliographystyle{plain}

\newpage ~ 
\end{document}